\documentclass[12pt]{article}

\usepackage[english]{babel}
\usepackage{graphicx}
\usepackage{subcaption}
\usepackage{authblk}
\usepackage{enumitem}
\usepackage{comment}

\newcommand{\R}{\mathbb{R}}

\newcommand{\C}{\mathbb{C}}

\usepackage{siunitx}

\usepackage[margin=2.5cm]{geometry}
 \usepackage{amsmath, amssymb}
\usepackage{amsthm}

\usepackage{xcolor}
\usepackage[colorlinks = true,
            linkcolor = blue,
            urlcolor  = blue,
            citecolor = blue,
            anchorcolor = blue]{hyperref}

\newcommand\extrafootertext[1]{%
    \bgroup
    \renewcommand\thefootnote{\fnsymbol{footnote}}%
    \renewcommand\thempfootnote{\fnsymbol{mpfootnote}}%
    \footnotetext[0]{#1}%
    \egroup
}

\usepackage{caption}
\usepackage{subcaption}

\usepackage{array,multirow}
\usepackage{hhline}

\begin{document}

\title{Estimating properties of a homogeneous bounded soil using machine learning models}

\author[1]{K. Kalimeris\thanks{kkalimeris@Academyofathens.gr}}
\affil[1]{Mathematics Research Center, Academy of Athens, Greece}
\author[2]{L. Mindrinos\thanks{leonidas.mindrinos@aua.gr}}
\affil[2]{Department of Natural Resources Development and Agricultural Engineering, Agricultural University of Athens, Greece}
 \author[3]{N. Pallikarakis\thanks{npall@central.ntua.gr}}
\affil[3]{Department of Mathematics, National Technical University of Athens, Greece}

\maketitle

\begin{abstract}

This work focuses on estimating soil properties from water moisture measurements. We consider simulated data generated by solving the initial-boundary value problem governing vertical infiltration in a homogeneous, bounded soil profile, with the usage of the Fokas method.
To address the parameter identification problem, which is formulated as a two-output regression task, we explore various machine learning models. The performance of each model is assessed under different data conditions: full, noisy, and limited. Overall, the prediction of diffusivity $D$ tends to be more accurate than that of hydraulic conductivity $K.$ Among the models considered, Support Vector Machines (SVMs) and Neural Networks (NNs) demonstrate the highest robustness, achieving near-perfect accuracy and minimal errors.

\textit{Keywords: Vertical infiltration; inverse problem; soil properties; Machine learning; multi-output regression.}

\end{abstract}

\extrafootertext{2020 Mathematics Subject Classification: 35G16; 35R30; 76S05; 86-10.}

\section{Introduction}\label{sec_intro}

Accurate estimation of soil properties, such as hydraulic conductivity $K$ and water diffusivity $D$, plays a crucial role in agricultural engineering \cite{raw82, tra86}. Traditional laboratory-based approaches, while reliable, are often time-consuming and require specific equipment. In recent years, data-driven techniques, particularly those from machine learning (ML), have emerged as powerful alternatives to predict complex material behavior with high efficiency and precision.

The inverse problem of determining soil properties from water content measurements at specific depths over a series of time steps is commonly formulated as a minimization problem involving a misfit function. This approach has been applied using deterministic methods to extract soil characteristics from field data, as demonstrated in several studies, see e.g. \cite{abbasi03, guel20, bour16, ritter03, si00, simu96}.  More recent research has shifted toward employing machine learning models for predicting soil properties. However, the majority of these studies emphasize to the recovery of soil chemical properties rather than physical characteristics using remote sensing data \cite{diaz22, for17, kha18}. Hyperspectral imaging data \cite{fou25}, digital mapping data \cite{zera19}, and electrical conductivity measurements \cite{mog19} are also used to estimate chemical properties using ML models. In \cite{pada19} convolutional neural networks (CNNs) were used to infer physicochemical and biological soil properties from spectroscopic data.

This study explores and compares the effectiveness of five well-known machine learning algorithms -- Support Vector Machines (SVM), Random Forests (RF), Extreme Gradient Boosting (XGBoost), Neural Networks (NN), and k-Nearest Neighbors (kNN) -- in estimating the two key soil properties. These algorithms were selected for their diverse modeling strategies and their success in regression tasks in different scientific disciplines. The models are trained and tested on simulated data obtained by numerically evaluating the integral representation of the solution of the partial differential equations (PDEs) governing the infiltration problem under consideration. More specifically, the dataset refers to a bounded homogeneous soil under the vertical infiltration process (flooding) and consists of measurements of water moisture at different depths over a time period for various pairs of conductivity and diffusivity. Each ML model undergoes a training-validation-testing pipeline to measure its prediction accuracy and generalization performance across the dataset.

Using simulated data with added noise to replicate field conditions displays both advantages and limitations in the context of estimating soil properties. On the positive side, simulated datasets allow for controlled experimentation by varying input parameters and generate large, diverse datasets that may be difficult or costly to obtain from field studies. Introducing noise into the simulation further enhances the realism of the data, helping to test model's stability. However, a key drawback lies in the potential discrepancy between simulated and real data, since they fail to capture the full complexity and heterogeneity of field conditions. Therefore, while simulated noisy data serve as a valuable tool for initial model development and testing, our aim is to examine the performance of the ML models with field data in future work.

The generation of simulated data is done as follows: We analytically treat the direct problem of computing water content of given soil properties using the Fokas method \cite{fokas97} for solving initial boundary value problems for linear PDEs.
In \cite{arg24}, two of the authors studied vertical infiltration in a homogeneous, bounded soil profile under flooding conditions, deriving the integral representation of the associated solution. This solution has been validated against well-established approximate analytical solutions and accurately models the described physical scenario. The synthetic data are obtained by the numerical computation of this integral representation, exploiting the favorable form of the relevant integrands. Finally, we add Gaussian noise to the test set to simulate realistic field measurements.

The motivation of this work lies in the effective generation of reliable train/test data, by evaluating the solution of the direct problem for several (not necessarily neighboring) points $(x,t)$,  via the above procedure. This abundance of data allows for a thorough comparison of the ML models. The objective of this research is not only to identify the most suitable algorithm for estimating soil properties, but also to provide insights into the performance and robustness of different ML techniques within a soil physics context. Through this comparative analysis, the study aims to contribute to the advancement of data-driven approaches for soil characterization. The effectiveness of the models is examined under the affection of the size of the dataset, the features and the added noise.

The structure of the paper is as follows. In \autoref{sec_problem}, we introduce the mathematical model for the vertical infiltration problem under flooding conditions, considering a homogeneous, bounded, one-dimensional domain with constant soil properties. We present the analytical solution to the resulting initial boundary value problem and describe the procedure for generating simulated data that serve as features for the machine learning model. The \autoref{sec_methods} outlines the fundamental principles of the machine learning methods employed in this study. 
The performance of the various machine learning models is evaluated in \autoref{sec_recon}. We begin by considering the full dataset, consisting of 2000 samples and 30 features. To ensure robust model assessment and prevent underfitting or overfitting, cross-validation is employed. Additionally, we perform feature selection and reduction to identify the features  that drive the model's performance, thereby enhancing interpretability in an explainable ML framework. Finally, we investigate the models' performance under conditions of noisy and limited data.

\section{Problem formulation}\label{sec_problem}

We consider the vertical infiltration problem in a homogeneous bounded soil profile with length $L>0.$ The soil is characterized by the hydraulic conductivity $K$ and the water diffusivity $D.$ We assume that the water is applied by flooding in the surface but the following analysis covers also the rainfall problem.

Mathematically, the aforementioned problem is formulated through the following initial boundary value problem for the water content $\theta :[0,L]\times \R_{+} \rightarrow \R_{+}$:
\begin{subequations}\label{bvp}
\begin{alignat}{3}
\frac{\partial \theta}{\partial t} + K \frac{\partial \theta}{\partial x} &= D \frac{\partial^2 \theta}{\partial x^2},  \quad && 0<x <L, \, t>0,  \label{bvp1}\\
\theta (x,0) &= \theta_0, \quad && 0 <x <L, \label{bvp2}\\ 
\theta (0,t) = \theta_1, \,\, \theta (L,t) &= \theta_0, \quad &&t>0. \label{bvp3}
\end{alignat}
\end{subequations}
 The initial water content $\theta_0$ and the water content at the surface  $\theta_1$ are constants. In \cite{arg24} an analytical solution of \eqref{bvp} was derived using the Fokas method \cite{fokas97}. The solution admits the form
 \begin{equation}\label{sol_ex1}
\begin{aligned}
\theta (x,t) = \theta_0 + (\theta_0 - \theta_1) \frac{2D}{\pi} e^{-\tfrac{K}{2D} (L-x)} \int_{C} \frac{1 - e^{-\omega(\lambda)t}}{\omega (\lambda) \Delta (\lambda)} \rho (\lambda) \sin \left( (L-x) \rho (\lambda)\right) d\lambda,
\end{aligned}
\end{equation}
where
\begin{equation}
\begin{aligned}
\rho (\lambda) &= \lambda + i \frac{K}{2D}, \\
\omega (\lambda) &=  D  \lambda^2 + i K \lambda, \\
\Delta (\lambda) &= e^{-i\lambda L}- e^{i \left(\lambda + i\tfrac{K}{D} \right) L}
\end{aligned}
\end{equation}
and the contour of integration is defined as
$$C = \big\{ \lambda \in \C: \operatorname{Re}(\omega (\lambda)) = 0, \ \operatorname{Im}(\lambda) > 0 \big\}.$$

To derive the solution \eqref{sol_ex1}, the central step involves the so-called global relation (GR), which is obtained using the finite interval Fourier transform. This relation connects specific (integral) transforms of both known and unknown boundary values. Then, a two-fold usage of this GR yields the desired result, namely the integral representation \eqref{sol_ex1}, which incorporates (transforms of) the initial and boundary data. First, inverting the GR yields an integral representation that involves both known and unknown boundary values; second, the symmetries of the GR allow the computation of the contribution of the unknown values to this integral representation of the solution. In \cite{arg24} there is extensive discussion on favorable contour deformation of $C$, which (i) retains the exponential decay of the integrand and (ii) avoids the contribution of possible singularities of the integrand, coming from the roots of the denominator. Therein, the illustration of the solution is performed, via numerical integration, for different setups for a very low computational cost without sacrificing the high accuracy.

The direct problem consists of finding $\theta,$ given the soil parameters, the length of the profile, and the initial and boundary conditions. We are interested in solving the corresponding inverse problem of recovering the soil parameters, given $\theta$ for some $x\in (0,L)$ and $t>0,$ as well as the length of the profile and the initial and boundary conditions. In this procedure, we train the ML models with data generated by the evaluation of the solution \eqref{sol_ex1} at a set of discrete points $(x_i,t_j)$, taking advantage of its computationally efficient features. Although the presence of non-zero conductivity $K$ complicates the inverse problem, it does not impact the generation of our training and testing datasets, which are constructed using the direct problem in the same manner as when $K = 0$.

\subsection{Simulated data}

We mimic measured data in the field by collecting water content values at few specific depth positions for different time steps. We set $x\in \{\tfrac{L}{4},\,\tfrac{L}{2},\,\tfrac{3L}{4} \}$ and $t =t_j, \, j=1,...,J.$ Thus, the data consists of 
\[
\Theta_{k,j} = \theta (x_k, t_j), \quad \mbox{for } k=1,2,3, \, j=1,...,J.
\]
In practice, the data is not perfect and might vary due to sensor calibration issues or weather conditions. We handle this problem by adding Gaussian noise to the simulated data
\begin{equation}\label{eq_error}
\Theta_\delta = \Theta + \epsilon,    
\end{equation}
where $\epsilon \sim \mathcal{N} (0, \sigma^2)$ follows a normal distribution with zero mean and standard deviation $\sigma = \delta \sigma_\Theta,$ where $\delta$ is the noise level and $\sigma_\Theta$ is the standard deviation of $\Theta.$ This way, the noise magnitude is proportional to the variability of the data.  Noise is introduced exclusively in the test set, ensuring that the model is trained solely on exact (noise-free) data and is exposed to noisy inputs only during evaluation. This setup poses a greater challenge for the model compared to scenarios where it is trained with noisy data, as it must generalize to distortions it has never encountered before.

\begin{figure}[t!]
    \centering
    \begin{minipage}{0.45\textwidth}
        \centering
        \includegraphics[width=\linewidth]{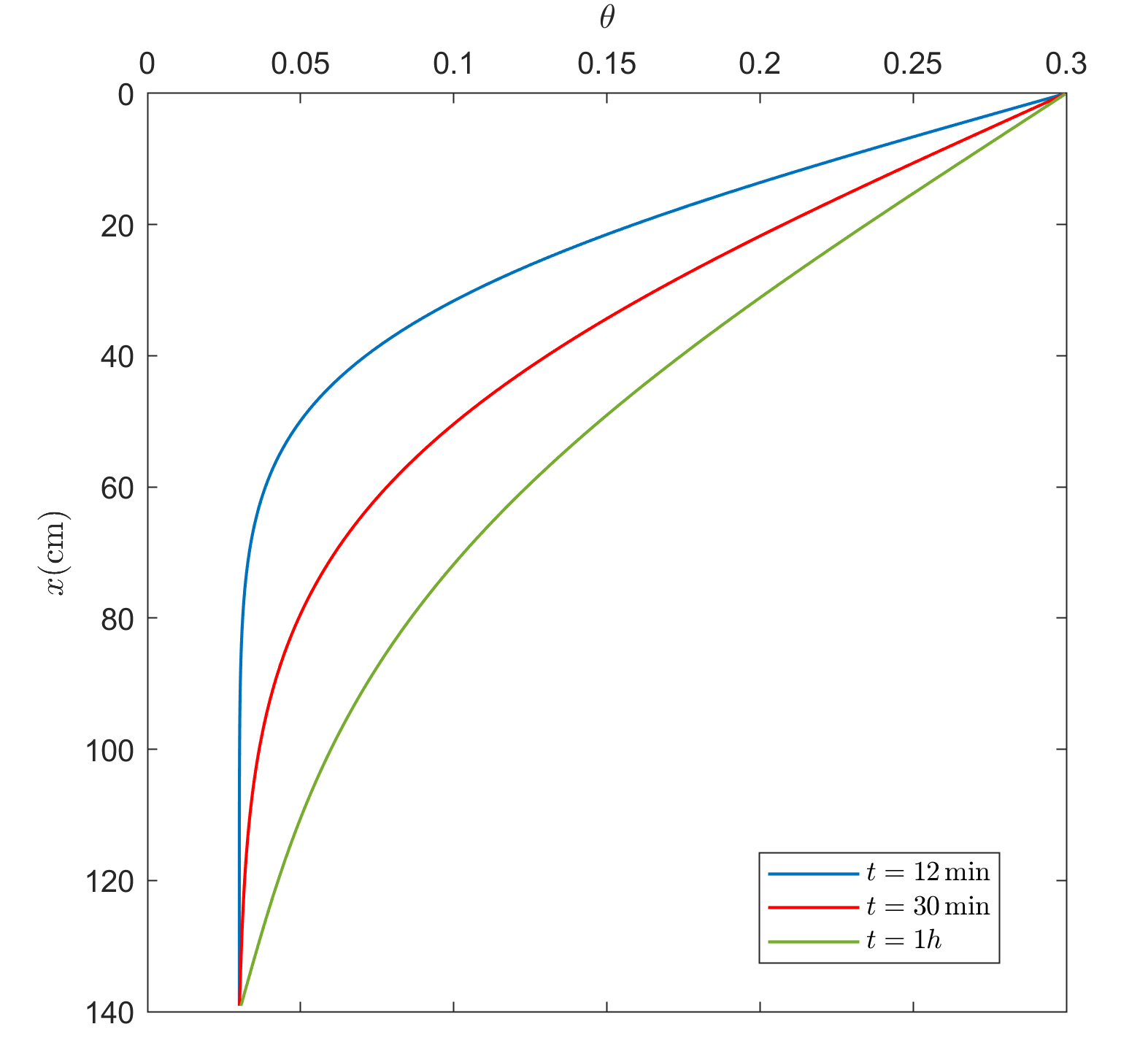}
        \label{fig:first}
    \end{minipage}
    \hfill
    \begin{minipage}{0.45\textwidth}
        \centering
        \includegraphics[width=\linewidth]{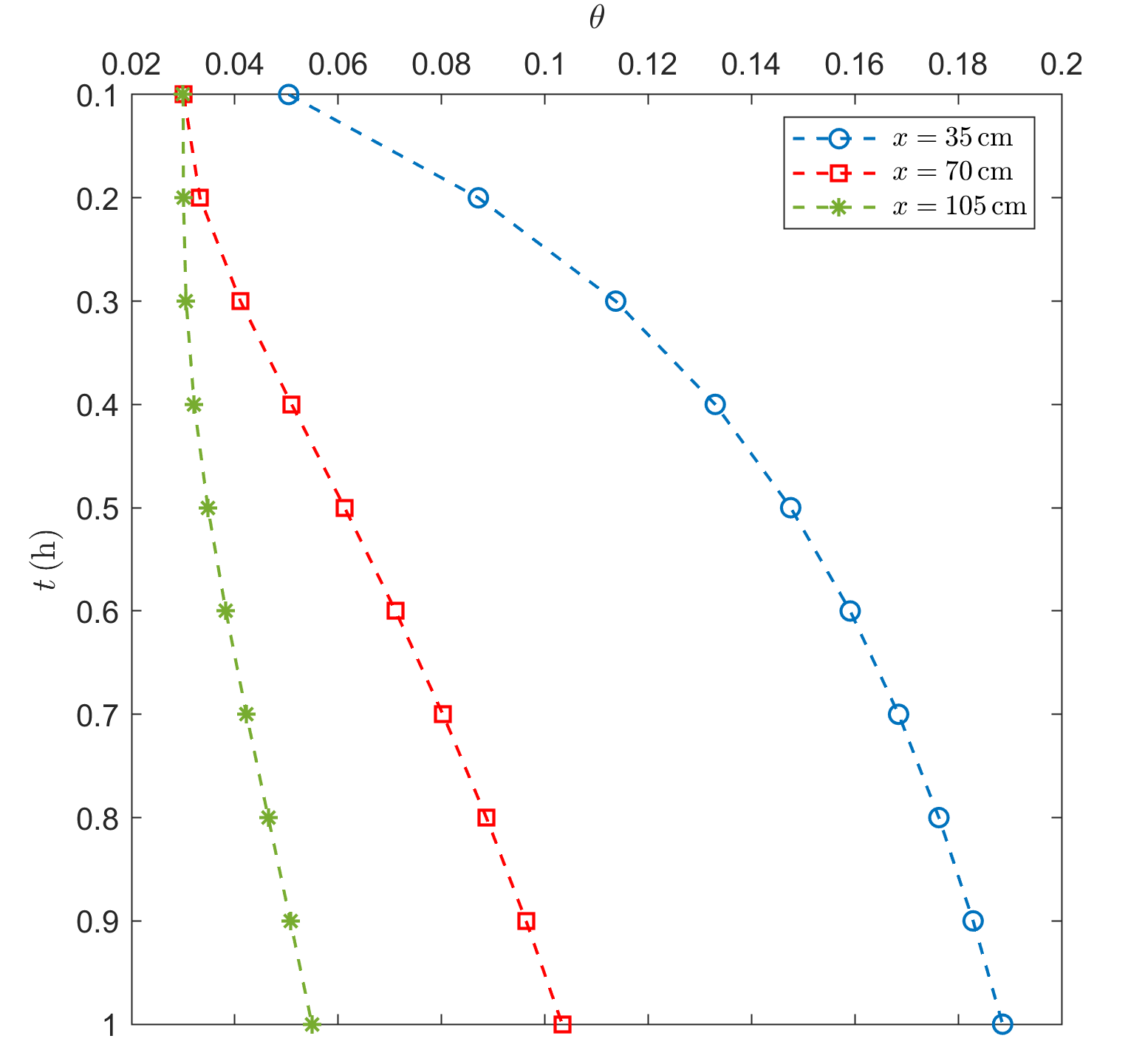}
        \label{fig:second}
    \end{minipage}
    \caption{Moisture profiles $\theta$ in soil for different infiltration times due to flooding (left). Moisture profiles at given depth positions with respect to time (right).}
    \label{fig1}
\end{figure}

We set the profile length $L=\SI{140}{\cm}$ and the initial and boundary conditions \cite{Arg97}
\[
\theta_1 = \SI{0.3}{cm^3 / cm^3}  \quad \mbox{and} \quad \theta_0 = \SI{0.03}{cm^3 / cm^3}.
\]
The soil parameters $K$ and $D$ are randomly generated using
\[
K = (3+3 \epsilon_K) \,\SI{}{cm/h} \quad \mbox{and} \quad D = (0.3 + 0.3 \epsilon_D) \,\SI{}{cm^2/sec},
\]
where $\epsilon_K, \, \epsilon_D \sim \mathcal{U} (0, 1)$ are uniformly distributed random numbers in the interval $[0,1].$ Thus, the parameters vary in the intervals $K \in [ 3,\,6] \,\SI{}{cm/h}$ and $D \in [1080,\, 2160] \,\SI{}{cm^2/h}.$ In \autoref{fig1} we present the distribution of the water content $\theta,$ given by \eqref{sol_ex1}, for $K = \SI{3.062}{cm/h}$ and $D =\SI{1911.6}{cm^2/h}.$ In the left picture, we present the water content $\theta$ at the time steps $t= \SI{12}{min},\, \SI{30}{min}$ and $\SI{60}{min}.$ We clearly observe that at $x=0,$ we get $\theta=\theta_1$ and as $x$ increases, the solution converges to $\theta_0.$ The simulated data - features used in the machine learning models are presented in the right picture. We specify three collecting data points at $x= L/4, \, L/2$ and $3L/4$ and calculate the moisture profile for $t_j=0.1 j,\, j=1,...,10$ (marked on the curves). Thus, for a given pair of $(K,D)$ we collect $\Theta \in \R^{3\times 10}.$ 

Initially, we consider 2000 randomly generated pairs of $(K,D)$ and we reshape the data $\Theta$ as
\[
\tilde{\Theta} = [\Theta_{1,1}, \Theta_{1,2},...,\Theta_{2,1},...,\Theta_{2,10},\Theta_{3,1}, ...,\Theta_{3,10}] \in \R^{1\times 30}.
\]

Thus, the input data are represented as $X\in\R^{2000\times 30}$, while the corresponding labels to be estimated are given by the pairs $(K,D)\in \R^{2000 \times 2}$. In addition, we will examine scenarios with limited data by reducing the number of data points or the number of time steps.

\section{Methods}\label{sec_methods}

In this section, we detail the methodology adopted in this study, beginning with a dimensionality reduction procedure applied as an initial preprocessing step to extract dominant patterns from high-dimensional, time-dependent data. The dataset comprises time series measurements, which often exhibit multicollinearity among features--a condition that undermines the assumptions of linear models and compromises their predictive power. Given these limitations, and recognizing the nonlinear and dynamic nature of soil-water interactions, we focus exclusively on nonlinear machine learning techniques. These models are better equipped to capture complex interactions and nonlinear dependencies that characterize the relationship between soil properties and observed data. Feature analysis techniques are also examined to identify a minimal yet informative subset of features that enables the development of efficient and accurate models. The section concludes with the performance metrics employed to evaluate their predictive accuracy.

\subsection{Dimensionality Reduction}\label{subsec_reduction}

Dimensionality reduction is essential in multi-output regression to improve model efficiency and generalization, especially for high-dimensional (i.e., $30$) feature data. Principal Component Analysis (PCA) \cite{pca} and Uniform Manifold Approximation and Projection (UMAP)  \cite{umap} are two complementary techniques for this task. PCA identifies the orthogonal directions of maximal variance and reveals linear relationships among variables, while UMAP uncovers nonlinear relationships, preserving both local and global structures. Together, these embeddings help us detect outliers, understand how smoothly each target varies across feature space, and decide whether to use simple or nonlinear models. In this work, dimensionality reduction serves a dual role: as a preprocessing step to visualize data structure, but also as a feature transformation strategy.

\subsection{Machine learning models}
For the reader's convenience and to keep the manuscript concise, we present only the main ideas of the ML models used in this study. We refer to the comprehensive bibliography, such as \cite{deep,elem} for a detailed review.

Support Vector Machines (SVMs) are supervised learning models that can be adapted for regression tasks, known as Support Vector Regression (SVR). The core idea of SVR is to find a function that approximates the data within a certain margin of tolerance, defined by the $\epsilon$ parameter.  Key hyperparameters in SVR include the kernel function, which transforms complex data patterns into a higher-dimensional space where they become more manageable for the model. The regularization parameter $c$ balances the trade-off between minimizing training error and maintaining model generalization. The kernel coefficient $\gamma$, relevant for kernels such as the Radial Basis Function (RBF), determines the extent of influence of individual training samples, with lower values implying broader influence and higher values resulting in more localized effects.

\begin{table}[h!]
\centering
\begin{tabular}{|l|l|}
\hline
\textbf{Component} & \textbf{Hyperparameter / Setting} \\
\hline
Kernel & Radial basis function \\ \hline
Regularization parameter $c$ & 1000 \\ \hline
Kernel coefficient $\gamma$ & 0.01 \\ \hline
Estimator $\epsilon$ & 0.01 \\ \hline
\end{tabular}
\caption{Hyperparameters and architecture of the support vector regressor.}
\label{table:svm_hyperparams}
\end{table}

Extreme Gradient Boosting (XGBoost) is an ML algorithm that builds an ensemble of decision trees through boosting to predict continuous values. Its most important hyperparameters include the number of estimators, which control the number of trees in the model; the maximum depth, which determines the maximum depth of each tree; the learning rate, which controls how much the model is updated during training; and regularization parameters, to add penalty weights to the loss function, helping to prevent overfitting by discouraging large coefficients. 

\begin{table}[h!]
\centering
\begin{tabular}{|l|l|}
\hline
\textbf{Component} & \textbf{Hyperparameter / Setting} \\
\hline
Loss function & Squared error \\ \hline
Estimators & 300 \\ \hline
Max depth  & 7 \\ \hline
Learning rate  & 0.2  \\ \hline
$L^2$ regularization parameter  & 100 \\ \hline
\end{tabular}
\caption{Hyperparameters and architecture of the XGB regressor.}
\label{table:xg_hyperparams}
\end{table}

Random Forest (RF) is another ensemble method based on decision trees. Key hyperparameters include the number of trees and max depth, like in XGBoost. Additionally, the minimum number of samples required to split an internal node (min sample split), which helps prevent the model from learning overly specific patterns; the minimum number of samples required to be at a leaf node (min sample leaf), which ensures that each leaf has enough samples to make reliable predictions; and the number of features to consider when looking for the best split (max features), which introduces randomness into the model.

\begin{table}[h!]
\centering
\begin{tabular}{|l|l|}
\hline
\textbf{Component} & \textbf{Hyperparameter / Setting} \\
\hline
Number of trees & 100 \\ \hline
Maximum depth & 20 \\ \hline
Min sample split & 2 \\ \hline
Min sample leaf &  1 \\ \hline
Number of features &  \texttt{Sqrt} \\ \hline
\end{tabular}
\caption{Hyperparameters and architecture of the random forest regressor.}
\label{table:rf_hyperparams}
\end{table}

Fully connected Neural Networks (NNs), or Multi-Layer Perceptrons (MLPs), are a class of deep learning models composed of multiple layers of neurons, where each neuron in one layer is connected to every neuron in the subsequent layer. These networks are specialized in learning complex patterns from data through backpropagation and optimization techniques. Main hyperparameters in MLPs include the number of layers and the number of nodes per layer, which define the network's capacity and complexity (i.e. depth). The activation function introduces non-linearity, enabling the network to model complex data relationships. Regularization techniques help prevent overfitting by penalizing large weights. The optimizer adjusts the weights during training to minimize the loss function. Early stopping monitors the validation loss during training and stops the process when performance cannot improve further, thereby preventing overfitting and reducing unnecessary computation. We refer the reader to the supplementary material of \cite{Pal2024} for a comprehensive presentation of the MLP algorithm.

\begin{table}[h!]
\centering
\begin{tabular}{|l|l|}
\hline
\textbf{Component} & \textbf{Hyperparameter / Setting} \\
\hline
Activation function & \texttt{LeakyReLU} with $\alpha=0.01$ \\ \hline
Regularization & $L^2$ with parameter $2\times10^{-4}$ \\
\hline
Layer 1 & Dense with 128 nodes  \\
\hline
Layer 2 & Dense with 64 nodes  \\
\hline
Layer 3 & Dense with 32 nodes  \\
\hline
Output layer & Dense with 2 nodes \\
\hline
Optimizer & Adam with learning rate 0.0001 \\
\hline
Loss function & Mean Squared Error (MSE) \\
\hline
Epochs & 2000 \\
\hline
Early stopping & Patience = 50 and $\Delta = 10^{-4}$ \\
\hline
\end{tabular}
\caption{Hyperparameters and architecture of the fully connected neural network model. }
\label{table:nn_hyperparams}
\end{table}

$k$-Nearest Neighbors (kNN) is a non-parametric, instance-based learning algorithm used for regression tasks. It predicts the target value of a data point by averaging the target values of its $k$ nearest neighbors in the feature space. The main hyperparameters include the number of neighbors which specifies the number of nearest neighbors to consider when making predictions; weight function which determines the weighting of the neighbors' contributions to the prediction and power parameter, to define the distance measure. When $p=1,$ it corresponds to the Manhattan distance and when $p=2,$ it corresponds to the Euclidean distance.

\begin{table}[h!]
\centering
\begin{tabular}{|l|l|}
\hline
\textbf{Component} & \textbf{Hyperparameter / Setting} \\
\hline
Number of nearest neighbors & 3 \\ \hline
Weight function & \texttt{distance} \\ \hline
Power parameter  & 2 \\ \hline
\end{tabular}
\caption{Hyperparameters and architecture of the $k-$nearest neighbors regressor.}
\label{table:knn_hyperparams}
\end{table}

Selecting appropriate hyperparameters for each ML model is often a challenging task, influenced by factors such as dataset size, the complexity of the relationships between input features and target variables, and the available computational resources. For smaller datasets, such as those used in this study, an effective strategy is to conduct a grid search over a predefined set of candidate hyperparameter values. Grid search systematically evaluates all possible combinations within the specified grid to identify the configuration that yields the best model performance. The optimal hyperparameters identified for each model in our analysis are summarized in \autoref{table:svm_hyperparams} through \autoref{table:knn_hyperparams}.

For each combination of parameters, the model is trained and evaluated using $5-$fold cross-validation to ensure robustness and avoid overfitting. We use \texttt{GridSearchCV} from \texttt{scikit-learn} to find the  best-performing model configuration based on the mean squared error (MSE).
The full dataset is randomly split, with 80\% used for training, 10\% for validation, and 10\% is left unseen for testing. Then, as a pre-processing step, we scale our data. Data pre-processing transforms raw data into a clean and structured format, enabling ML algorithms to perform more effectively. We compute normalization parameters only on the training set and we apply  that same transformation to the validation and test sets. This is performed before training the ML models, to avoid any data leakage.  For the needs of our study, we used \texttt{StandardScaler} from \texttt{scikit-learn} to ensure that every feature has zero mean and standard deviation one.

\subsection{Cross Validation}

Cross-validation is a fundamental technique in machine learning used to assess a model’s ability to generalize to unseen data. By partitioning the dataset into multiple train-test splits, typically through $k-$fold or leave-one-out strategies, it allows for repeated training and evaluation of the model across different subsets. A common choice, used also in this work, is 5-fold cross-validation, where the data is divided into five parts, with each part serving as the test set once while the remaining four are used for training. This process not only provides a more robust estimate of predictive performance but also helps detect issues such as overfitting, where a model learns the training data too closely, or underfitting, where it fails to capture the underlying patterns. As a result, cross-validation is essential for model selection, hyperparameter tuning, and ensuring the reliability of results, especially when working with limited data.

\subsection{Feature Analysis }

Feature reduction, selection and transformation are crucial steps in developing effective machine learning models, particularly when working with structured datasets, as in our case where we trained models using 30 measurements collected at three depths over ten time steps.  In domains where features are derived from physical or theoretical models and span multiple time steps, multicollinearity is commonly encountered, as illustrated in Figure \ref{fig1}. To better understand the contribution of each input and enhance model interpretability, we implement a structured explainable machine learning workflow that includes: (a) feature reduction, (b) feature selection, and (c) feature transformation. 

In our study, identifying the most and least important features corresponds to selecting a subset of moisture profiles—defined by specific depths and time intervals—that are most predictive of hydraulic conductivity ($K$) and water diffusivity ($D$). Beyond improving model interpretability and computational efficiency, this process holds practical relevance for agricultural engineering. By isolating the most informative measurements, it provides guidance for optimizing sensor placement and data acquisition strategies in real-world agricultural applications, thereby supporting more accurate and efficient soil moisture monitoring.

\subsubsection{Feature reduction - limited vertical resolution}

Feature reduction involves simplifying the input space by evaluating whether fewer measurements, such as data from a single depth, can achieve comparable predictive performance. This helps determine the minimal vertical resolution necessary for accurate modeling. By examining performance degradation with reduced input granularity, we identify which spatial or temporal dimensions may be redundant or less informative for the target predictions of $K$ and $D$.

\subsubsection{Feature selection - feature importance}

Feature selection aims to find the most influential variables from the full set of inputs. We apply a model-agnostic perturbation algorithm, where each feature is independently shuffled to break its relationship with the target, and the resulting change in model error is recorded. A larger increase in mean squared error signifies greater importance. These importance scores are then normalized to a scale between 0 and 1 for comparison. This algorithm was developed for NNs \cite{7966442}, which has also been effectively applied to other ML models for inverse problems, as demonstrated in \cite{PN}. 

\subsubsection{Feature transformation - dimensionality reduction}

To explore the structure of the input space and identify latent relationships, we apply dimensionality reduction techniques, including PCA and UMAP (see also \autoref{subsec_reduction}). These transformations compress the original 30-dimensional dataset into two dimensions, allowing us to visualize whether predictive information aligns with major variance directions (PCA) or with more complex nonlinear manifolds (UMAP). This step offers insights into the geometry of the data and whether effective modeling can be achieved through lower-dimensional representations.

\subsection{Metrics}

Let $y_i$ and $\tilde{y}_i,$ for $j=1,...,n$ denote the actual and the predicted values, respectively. The following metrics will be used to test the performance of the machine learning models:

 Mean absolute error:
\[
\mbox{MAE} = \frac{1}{n} \sum_{j=1}^n |y_i - \tilde{y}_i|
\]

 Mean squared error:
\[
\mbox{MSE} = \frac{1}{n} \sum_{j=1}^n (y_i - \tilde{y}_i )^2
\]
 
 $R^2$ score: 
\[
R^2 = 1- \frac{\displaystyle\sum_{j=1}^n (y_i - \tilde{y}_i)^2 }{\displaystyle\sum_{j=1}^n (y_i - \bar{y})^2},
\]
for the mean value $\bar{y}$ of the actual values. The $R^2$ score varies in the interval $(-\infty,1]$ with one meaning perfect prediction.

\section{Results}\label{sec_recon}

In this section, tables and plots will be used to visualize the performance of each model for both outputs. Especially, we use regression and residual plots.  A regression plot  shows the predicted values against the actual values.  The closer the points are to the line $y=x$ (called Perfect Fit) the better the predictions and a tight clustering around the diagonal line indicates high accuracy. A residual plot shows the residuals (errors) on the $y-$axis and the predicted values on the $x-$axis. Ideally, residuals should be randomly scattered around zero with no clear pattern. This suggests consistency and it is called $``$Homoscedasticity$"$ \cite{kutner2005applied}.

\subsection{Full data}

We begin by considering the full dataset, namely 2000 rows and 30 features - water content measurements. As an initial step, we apply both PCA and UMAP to project the high-dimensional feature space (i.e., 30 dimensions) into two dimensions. This dimensionality reduction serves as a preprocessing visualization technique to help us explore and understand the underlying structure of the data.

\begin{figure}[t!]
    \centering
        \includegraphics[width=\linewidth]{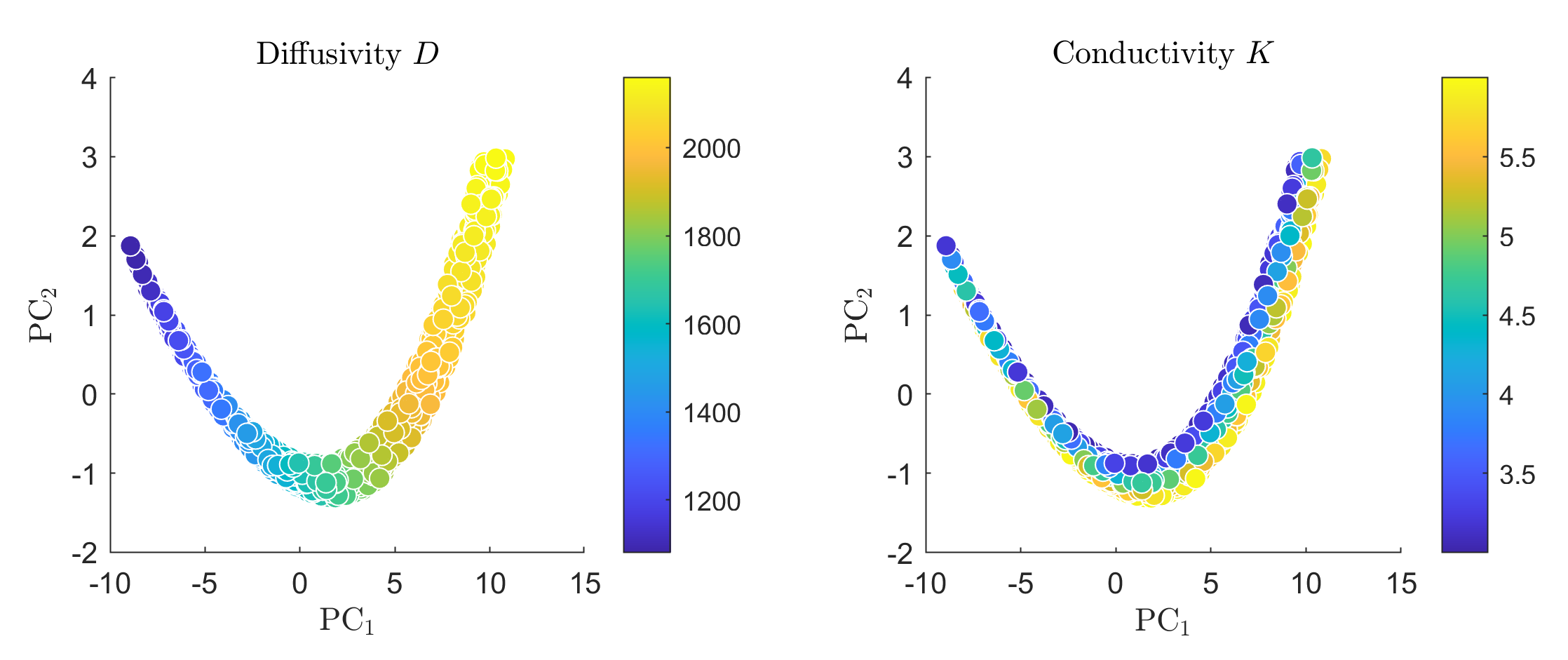}
    \caption{PCA dimension reduction for $D$ (left) and $K$ (right).}
    \label{figPCA}
\end{figure}

\begin{figure}[t!]
    \centering
        \includegraphics[width=\linewidth]{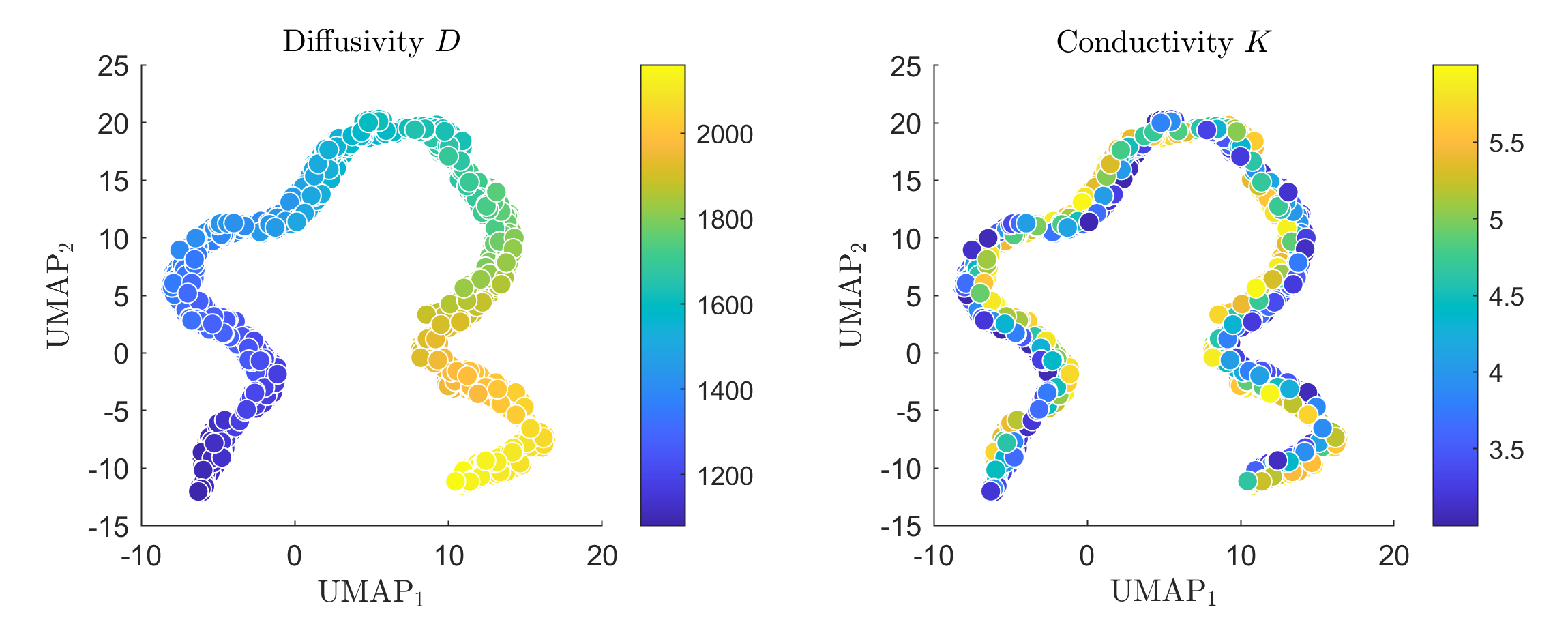}
    \caption{UMAP dimension reduction for $D$ (left) and $K$ (right).}
    \label{figUMAP}
\end{figure}

Both dimension reduction methods are unsupervised, which means that the projection into the lower dimension is performed without including any information about the target ($D$ and $K$) values. The 2D PCA projections are shown in \autoref{figPCA} and the corresponding 2D UMAP projections in \autoref{figUMAP}. The resulting scatter plots demonstrate a clear and smooth gradient of diffusivity $D$ in both PCA and UMAP embeddings, indicating a relatively straightforward relationship between the features and $D.$ In contrast, the distribution of conductivity $K$ appears less regular: in PCA space, lower $K$ values tend to cluster in the upper region of the PC$_1$–PC$_2$ plane, while in UMAP space, $K$ is more diffusely distributed, suggesting a more complex and nonlinear dependency on the input features. As we illustrate in the following, this increased complexity contributes to the relatively lower accuracy observed when predicting $K,$ particularly under noisy conditions, an outcome consistent with the insights from this dimensionality reduction analysis.

Next, we proceed with the results of the ML models, using the full dataset. In the test set, for predicting $D,$ SVM and NN achieved perfect $R^2$ scores (1.0000) indicating excellent fit but NN had much lower errors (MSE = 0.5293, MAE = 0.5640), suggesting better precision than SVM, see \autoref{table_metric_exactD}. XGBoost, RF, and kNN showed also perfect $R^2$ scores,  but higher errors, especially for RF and kNN, with MSE $> 14$ and MAE $> 2.7$ with kNN having the highest MSE (15.0795), indicating lower reliability for estimating $D.$

\begin{table}[t!]
\begin{center}
\begin{tabular}{| c | c  | c  | c  | c  | c| } 
 \hline
Metric & SVM & XGBoost & RF & NN & kNN  
\\ \hline 
$R^2$ score & 1.0000 & 0.9999 & 0.9999 & 1.0000 & 0.9998
\\
MSE &   2.9787 & 9.0156 & 14.4745 &  0.5293 &  15.0795
\\
MAE &  1.4625 & 2.2871 & 2.7821 & 0.5640 &  2.8197
 \\ \hline
\end{tabular}
\caption{Metrics for predicting the diffusivity $D$ in the test set using exact data.}\label{table_metric_exactD}
\end{center}
\end{table}

The performance of SVM and NN is again excellent for $K$ with the highest $R^2$ scores and almost zero MSE, as we see in \autoref{table_metric_exactK}. XGBoost, RF, and kNN still have good $R^2$ scores $(> 0.96),$ but with noticeably higher MAE (up to 0.1206 for RF). To summarize, SVM and NN consistently perform best across both parameters, while RF and kNN tend to show higher errors, especially in predicting diffusivity. XGBoost performs moderately well, with good but relatively higher errors.

\begin{table}[t!]
\begin{center}
\begin{tabular}{| c | c  | c  | c  | c  | c| } 
 \hline
Metric & SVM & XGBoost & RF & NN & kNN  
\\ \hline 
$R^2$ score & 1.0000 & 0.9710 & 0.9612 & 0.9999 & 0.9687
\\
MSE &   0.0000 & 0.0212 & 0.0284 &  0.0001 &  0.0229
\\
MAE &  0.0042 & 0.1044 & 0.1206 & 0.0064 &  0.1114
 \\ \hline
\end{tabular}
\caption{Metrics for predicting the conductivity $K$ in the test set using exact data.}\label{table_metric_exactK}
\end{center}
\end{table}

As revealed by our preprocessing analysis, conductivity $K$ appears more challenging to predict accurately than $D$, as shown in \autoref{fig_regre_exact}. All models perform excellent for $D$, yielding a perfect fit, while XGBoost, RF, and kNN show some variation for $K$. 

Regarding the residual plots (see \autoref{fig_res_exact}), we observe that, in all cases, the residuals are randomly distributed around zero without any clear curvature. The SVM model performs the best, with residuals ranging from $[-2.5,\, 2.5]$ for diffusivity and $[-0.01,\, 0.01]$ for conductivity. The NN also performs well, while kNN exhibits the largest residuals for both outputs.


In \autoref{table_reconD} and \autoref{table_reconK}, we compare the predicted values of all ML models against the actual values for both parameters, for five randomly selected samples. All models exhibit high prediction accuracy, with predicted values closely matching the actual ones. For diffusivity $D$ (see \autoref{table_reconD}), the Neural Network showed the lowest MAE (0.53), followed by XGBoost and SVM. The kNN model exhibited the highest variability. For conductivity $K$, SVM and NN achieved the lowest mean absolute errors (0.004), indicating excellent performance, while XGBoost had the highest average error in that subset (see  \autoref{table_reconK}).

\begin{table}[t!]
\begin{center}
\begin{tabular}{| c || c  | c  | c  | c  | c| } 
 \hline
Actual & SVM & XGBoost & RF & NN & kNN  
\\ \hline 
1684.59  &  1683.87  & 1684.87 & 1682.27 & 1685.58 & 1685.83 \\
1869.99  &  1869.60  & 1867.21 & 1870.23 & 1870.85 & 1868.19 \\
2117.94  &  2121.01  & 2116.82 & 2121.32 & 2118.22 & 2117.13 \\
1735.98  &  1737.62  & 1737.57 & 1739.88 & 1736.25 & 1744.83 \\
1382.93  &  1384.57  & 1384.15 & 1383.48 & 1382.67 & 1379.06 
 \\ \hline\hline
 MAE & 1.49 & 1.40 & 2.08 & 0.53 & 3.31
 \\ \hline
\end{tabular}
\caption{Randomly selected predicted values of the diffusivity $D$ in the test set for exact data.}\label{table_reconD}
\end{center}
\end{table}

\begin{table}[t!]
\begin{center}
\begin{tabular}{| c || c  | c  | c  | c  | c| } 
 \hline
Actual & SVM & XGBoost & RF & NN & kNN  
\\ \hline 
4.323  &  4.320  & 4.211 & 4.393 & 4.323 & 4.322 \\
4.596  &  4.602  & 4.648 & 4.606 & 4.595 & 4.651 \\
5.472  &  5.475  & 5.242 & 5.332 & 5.460 & 5.435 \\
5.305  &  5.302  & 5.409 & 5.174 & 5.302 & 5.053 \\
4.092  &  4.088  & 3.919 & 4.025 & 4.089 & 4.173 
\\ \hline\hline
 MAE & 0.004 & 0.134 & 0.084 & 0.004 & 0.085
 \\ \hline
\end{tabular}
\caption{Randomly selected predicted values of the conductivity $K$ in the test set for exact data.}\label{table_reconK}
\end{center}
\end{table}

\begin{figure}[t!]
    \centering
    \begin{minipage}{0.45\textwidth}
        \centering
        \includegraphics[width=\linewidth]{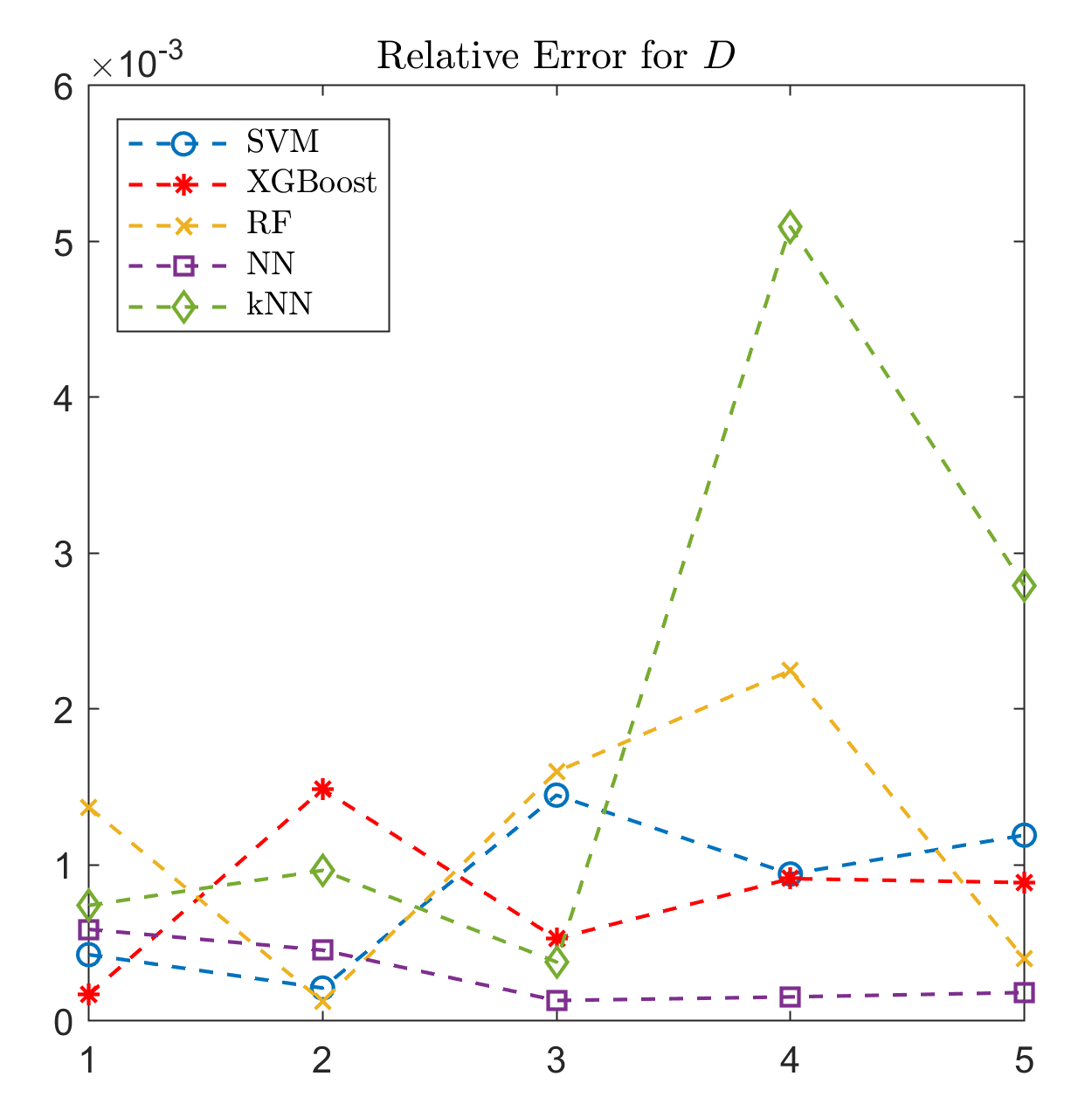}
        \label{fig:first1}
    \end{minipage}
    \hfill
    \begin{minipage}{0.47\textwidth}
        \centering
        \includegraphics[width=\linewidth]{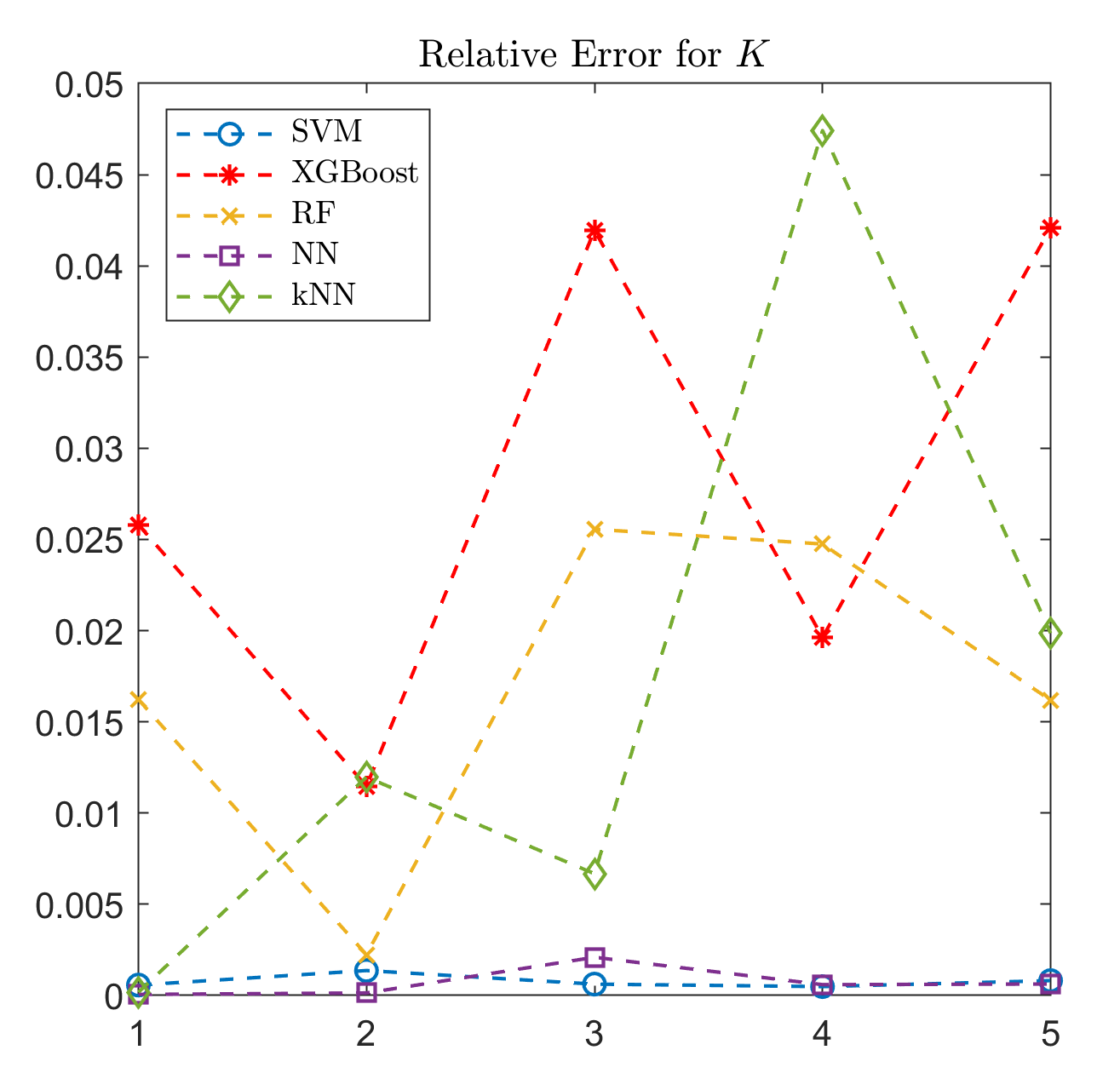}
        \label{fig:second2}
    \end{minipage}
    \caption{Relative errors between the actual and predicted values of  $D$ (left), as presented in \autoref{table_reconD}, and of $K$ (right), as shown in \autoref{table_reconK}. The indices on the $x$-axis represent the row numbers of the respective tables.}
    \label{fig_relative}
\end{figure}

In \autoref{fig_relative} we present the relative errors
\[
\mbox{RE} = \frac{| y_i - \tilde y_i|}{y_i}, \quad i=1,...,5,
\]
between the predicted $\tilde y_i$ and the actual values $y_i$ for the five selected pairs presented in \autoref{table_reconD} and \autoref{table_reconK}.  
These results suggest again that for this specific subset, NN offers the most consistent accuracy across both parameters. SVM and XGBoost perform similarly well for $D$ whereas for predicting $K,$ SVM tracks also closely the actual values.

\begin{figure}
\centering
  \includegraphics[width=0.6\textheight]{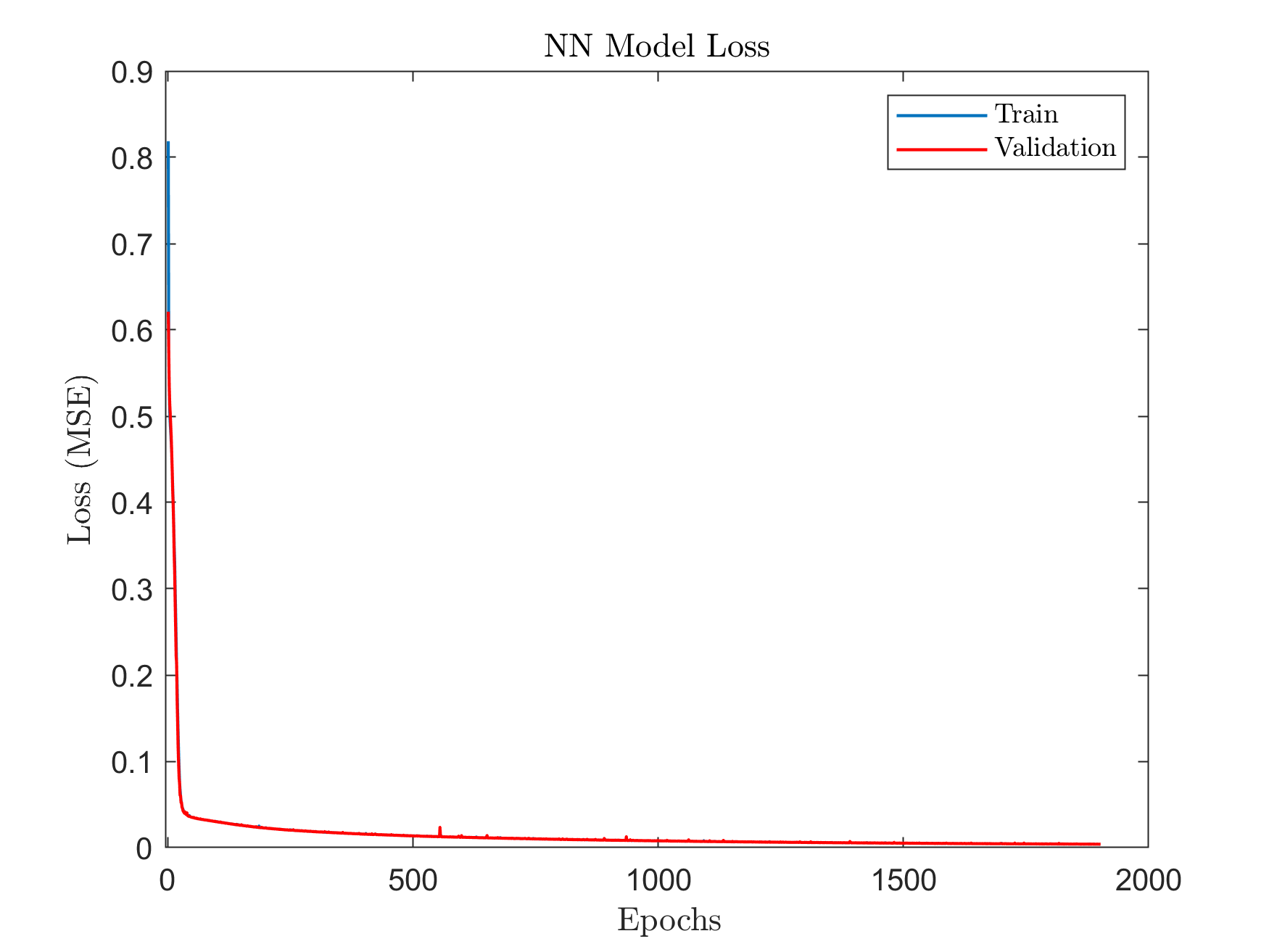}
\caption{The loss function of the neural network model for exact data. }
\label{fig_hist_ex}
\end{figure}

\begin{figure}[ht!]
\centering
\begin{subfigure}{.5\textwidth}
  \centering
  \includegraphics[height=0.9\textheight]{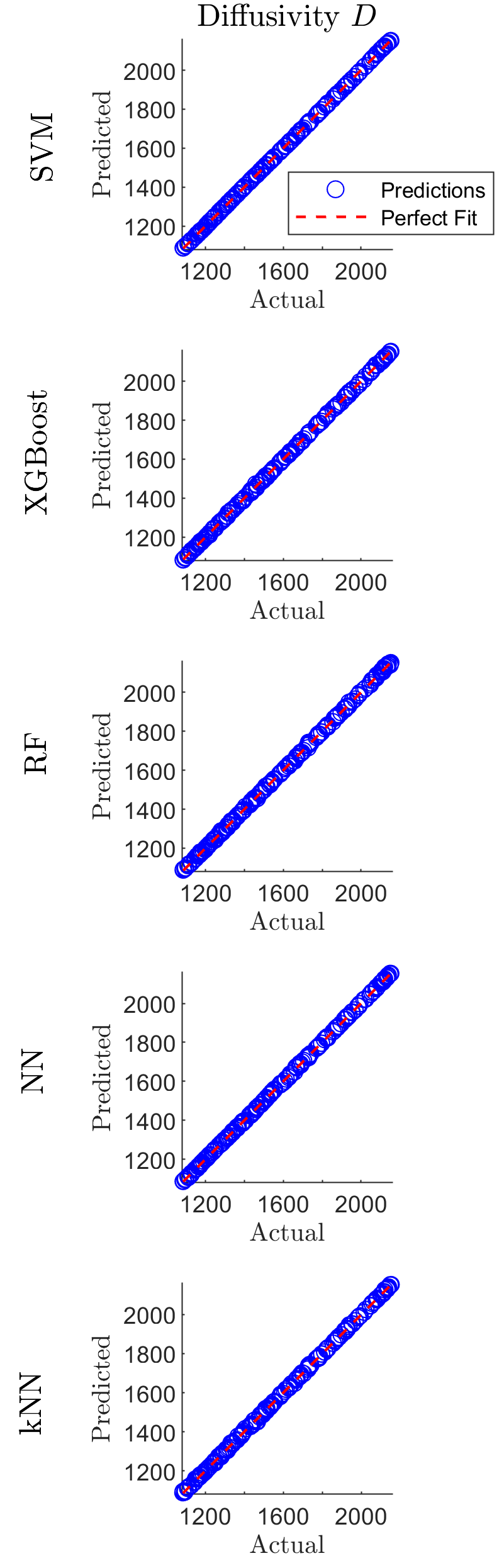}
\end{subfigure}%
\hspace{-1.5cm}
\begin{subfigure}{.5\textwidth}
  \centering
  \includegraphics[height=0.9\textheight]{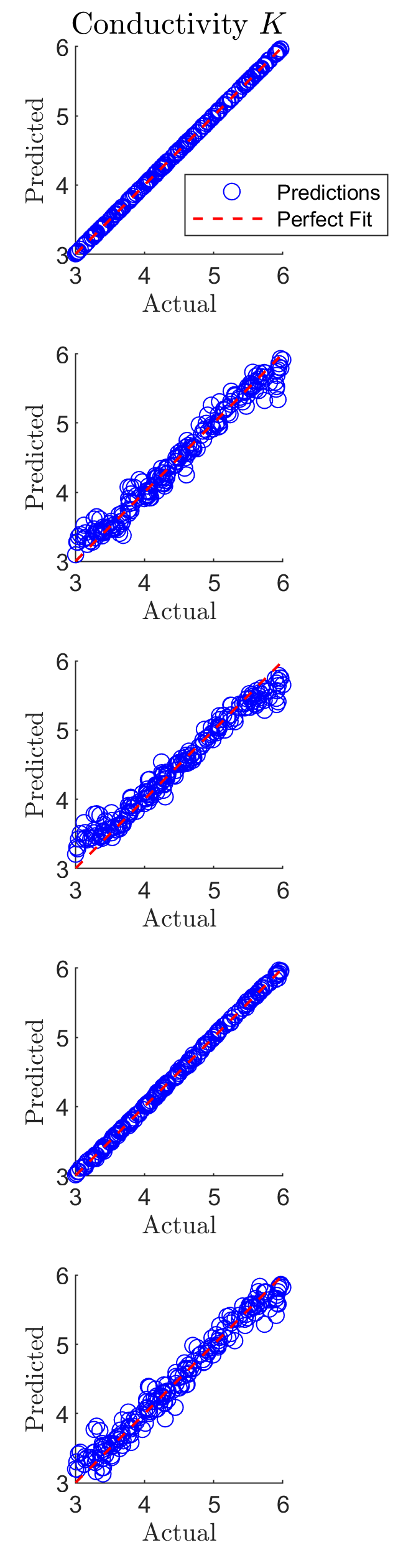}
\end{subfigure}
\caption{Regression plots of $D \,[\SI{}{cm^2/h}]$ (left) and $K \,[\SI{}{cm/h}]$ (right) in the test set for exact data.}
\label{fig_regre_exact}
\end{figure}

\begin{figure}[ht!]
\centering
\begin{subfigure}{.5\textwidth}
  \centering
  \includegraphics[height=0.9\textheight]{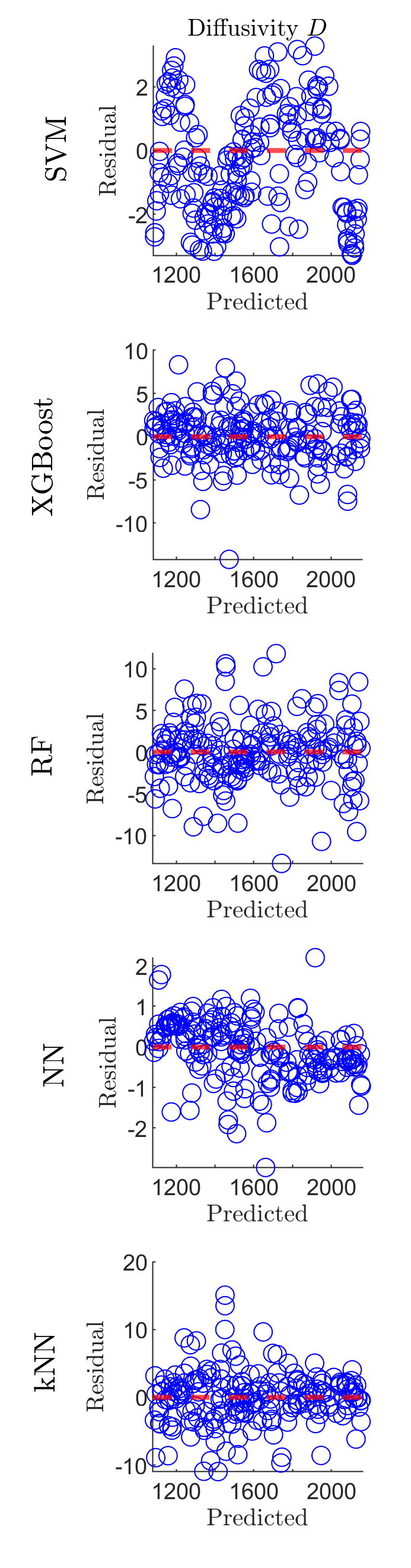}
\end{subfigure}%
\hspace{-1.5cm}
\begin{subfigure}{.5\textwidth}
  \centering
  \includegraphics[height=0.9\textheight]{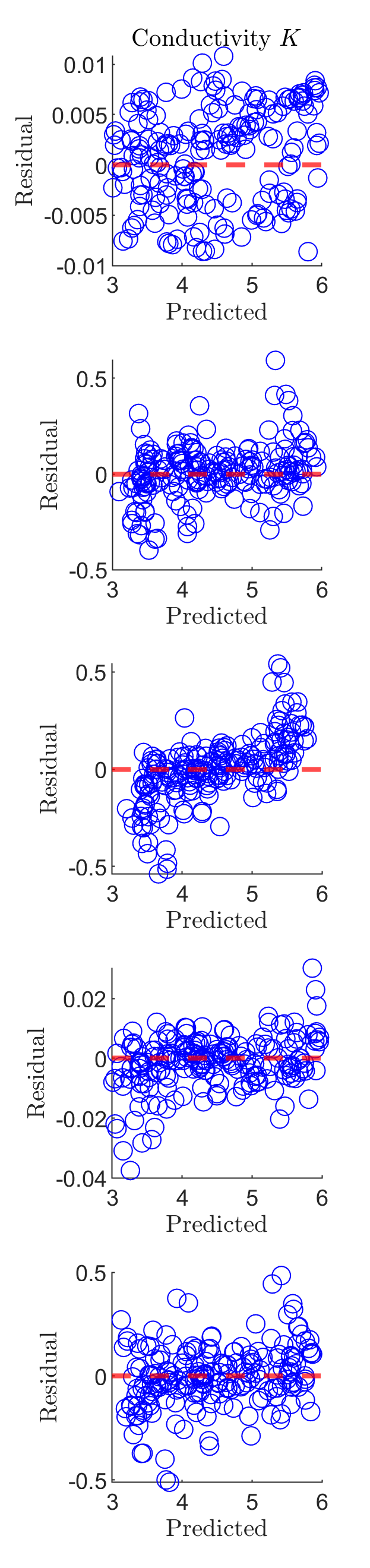}
\end{subfigure}
\caption{Residual plots of $D \,[\SI{}{cm^2/h}]$ (left) and $K \,[\SI{}{cm/h}]$ (right) in the test set for exact data.}
\label{fig_res_exact}
\end{figure}

The loss plot in \autoref{fig_hist_ex} represents the training and validation MSE of the neural network model over 2000 epochs. The model demonstrates a steep drop in both training and validation losses early in training, with MSE values falling sharply in the initial few hundred epochs. After that, the losses steadily converge towards near-zero values and remain very low throughout the remainder of the training process. The model does not reach 2000 epochs because of the stopping criterion, as previously discussed.

\subsection{Cross-validation}

In this work, we use $CV=5$ meaning a $5-$fold cross-validation where the data is divided into 5 equal parts. Each fold serves as the test set once while the remaining four are used for training.  The \(R^2\) score is averaged over the 5 iterations and its standard deviation is presented in \autoref{table_crossD} and \autoref{table_crossK} for the diffusivity and the conductivity, respectively.

The cross-validation results summarized in \autoref{table_crossD}  demonstrate outstanding performance across all five ML models for predicting \( D \). Each model achieves an exceptionally high score. Moreover, the standard deviation for all models is zero, indicating highly consistent performance across all five cross-validation folds. This suggests that the dataset and all models are able to capture the underlying patterns with near-perfect accuracy and stability.

\begin{table}[t!]
\begin{center}
\begin{tabular}{| c | c  | c  | c  | c  | c| } 
 \hline
 & SVM & XGBoost & RF & NN & kNN  
\\ \hline 
Mean  & 1.0000 & 0.9999 & 0.9998 &  1.0000 & 0.9998
\\
StDev. &  0.0000 & 0.0000 & 0.0000 & 0.0000 &  0.0000
 \\ \hline
\end{tabular}
\caption{Cross validation (CV$=5$) $R^2$ results for predicting the diffusivity $D$ using exact data.}\label{table_crossD}
\end{center}
\end{table}

In \autoref{table_crossK} the results for predicting the conductivity \( K \) show slightly more variability in model performance. While the SVM and NN still achieve near-perfect accuracy, other models show reduced performance, with mean \( R^2 \) scores around 0.96. Additionally, their standard deviations are noticeably higher (e.g., 0.0059 for XGBoost and 0.0057 for RF), indicating more variation across the cross-validation folds. This suggests that the conductivity prediction task is slightly more challenging than the diffusivity case, leading to greater sensitivity in some models. Nevertheless, all models still perform well, but the comparison highlights the superior consistency and accuracy of SVM and NN across both regression tasks.

\begin{table}[t!]
\begin{center}
\begin{tabular}{| c | c  | c  | c  | c  | c| } 
 \hline
 & SVM & XGBoost & RF & NN & kNN  
\\ \hline 
Mean  & 1.0000 & 0.9604 & 0.9539 &  0.9999 & 0.9623
\\
StDev. &  0.0000 & 0.0059 & 0.0057 & 0.0001 &  0.0038
 \\ \hline
\end{tabular}
\caption{Cross validation (CV$=5$) $R^2$ results for predicting the conductivity $K$ using exact data.}\label{table_crossK}
\end{center}
\end{table}

\subsection{Feature Analysis}

This section presents the results of three techniques that reduce the number of features, either by selecting a subset or by transforming them through projection.

\subsubsection{Limited vertical resolution}

We assess the performance of the machine learning models when progressively excluding the deepest measurement location, which corresponds to a reduction of 10 features at each step. Accordingly, we present results for models trained with 30, 20, and 10 features.

\autoref{fig_featD} and \autoref{fig_featK} illustrate how the key performance metrics change as the number of input features is reduced from 30 to 20 and then to 10 when predicting $D$ and $K,$ respectively. Among the models, SVM and NN demonstrate the highest resilience to feature reduction, maintaining consistently high \( R^2 \) scores and exhibiting only slight increases in MSE and MAE, even with just 10 features. XGBoost performs optimally with 20 features, achieving its highest \( R^2 \) and lowest error metrics when predicting \( K \), but its performance degrades as fewer features are used, particularly for predicting \( D \). RF shows stable performance between 20 and 10 features when predicting \( D \), but a noticeable drop when predicting \( K \) with fewer features. Interestingly, kNN is the only model that improves across all metrics as the number of features decreases from 30 to 10, specifically in the case of predicting \( K \).

In conclusion, feature constraints have minimal impact on the prediction of $D$, with only a slight increase in error observed. In contrast, the prediction of $K$ is more sensitive to feature reduction, depending on the model used. SVM and NN once again emerge as the most robust models, consistently maintaining high accuracy across different feature sets.

\subsubsection{Feature Importance}

\begin{figure}
\centering
  \includegraphics[height=0.9\textheight]{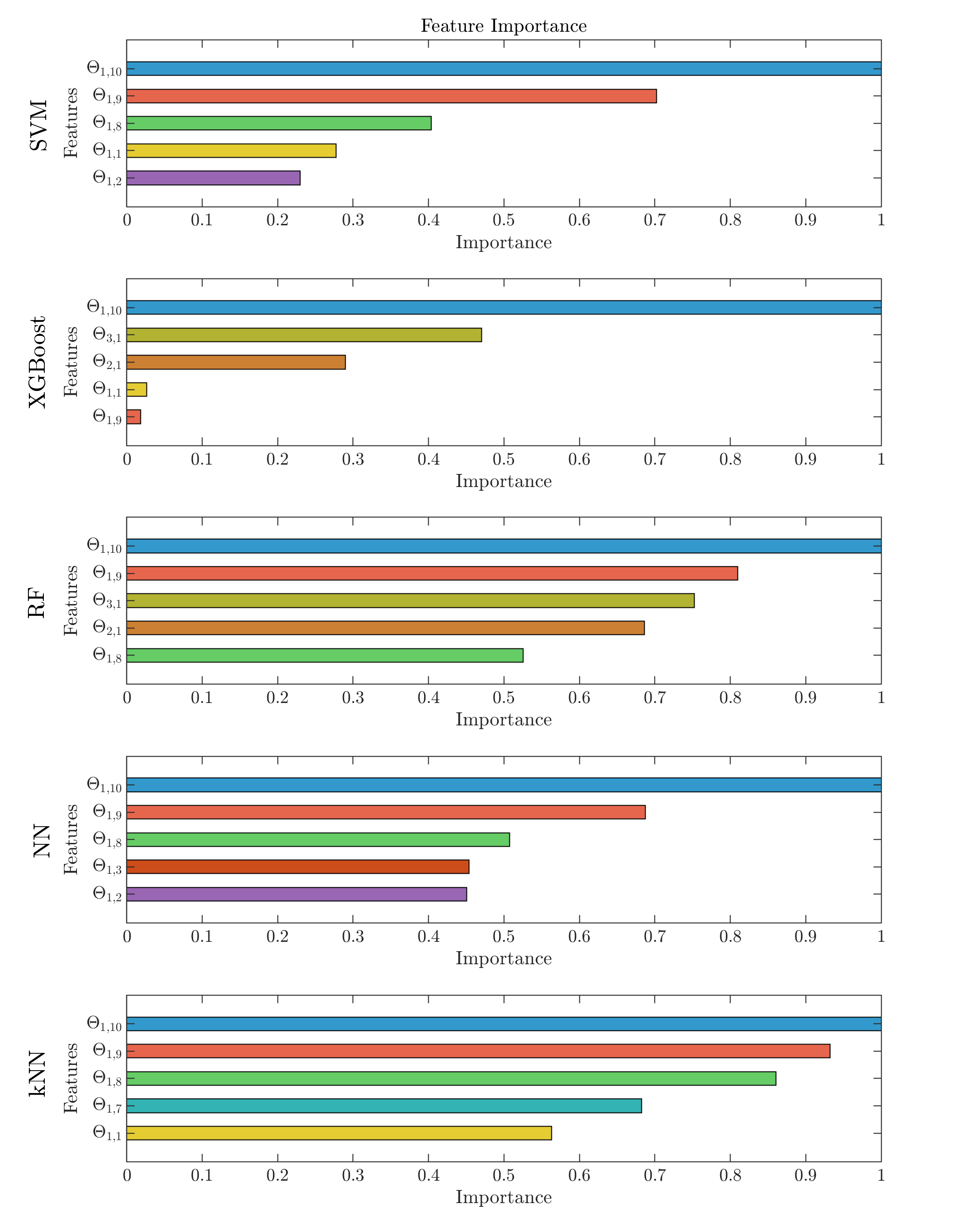} 
\caption{The five most important features identified for each model.}
\label{fig_importance}
\end{figure}

The feature importance analysis across all models highlights a clear dominance of the feature - measured data $\Theta_{1,10}$, corresponding to data collected at position $x = L/4\,\text{cm}$ and time $t = 1\, \text{h}$, which consistently holds the highest normalized importance score of 1.000. This indicates that it is the most influential input in terms of predictive power, with a significant margin over the other features for the SVM, XGBoost, and Neural Network models (see \autoref{fig_importance}). The next most important feature is $\Theta_{1,9}$, which ranks second in all models except XGBoost, where it appears in fifth place. Similarly, feature $\Theta_{1,8}$ is among the top five for most models, ranking third for SVM, NN, and kNN, though it is less important for XGBoost. Despite the effects of multicollinearity—which can distort importance scores among correlated features—the systematic agreement of these top-ranked features across all models suggests their robustness as key predictors.

Overall, the models tend to rely heavily on a small subset of features, particularly those measured closer to the surface ($x = 35\, \text{cm}$) and those recorded deeper at earlier times, which likely capture more distinct or informative dynamics relevant to the output prediction. This is a key observation with potential applications in soil data collection. 


\autoref{table_metric_exactDim} shows the performance and relative changes (in parentheses) in predicting diffusivity $D$ when using only the five most important features, as compared to the full feature set.  In terms of $R^2$ scores, all models maintained near-perfect predictive performance, with no significant changes.  However, relative changes in MSE and MAE reveal more variation. The SVM model exhibited the smallest increase in error suggesting minimal performance degradation. XGBoost has moderate increases in both MSE ($-22.0\%$) and MAE ($-12.2\%$), implying that it may have benefited slightly from feature reduction. On the other hand, RF and kNN suffered substantial increases in both error metrics. This suggests that these models are more sensitive to input dimensionality and rely more heavily on a broader set of features for accurate predictions.

\begin{table}[t!]
\begin{center}
\scalebox{0.85}{
\begin{tabular}{| c | c  | c  | c  | c  | c| } 
 \hline
Metric & SVM & XGBoost & RF & NN & kNN  
\\ \hline 
$R^2$ score & 1.0000 (0\%) & 0.9999 (0\%) & 0.9997 $(-0.02\%)$ & 1.0000 $(0\%)$ & 0.9998 (0\%)
\\
MSE &   2.9248 $(-1.8\%)$ & 7.0271 $(-22.0\%)$ & 26.0184 (+79.7\%) &  0.7718 $(+45.8\%)$ &  24.2639 (+61\%)
\\
MAE &  1.4709 (+0.6\%) & 2.0080
 $(-12.2\%)$ & 3.7125 (+33.4\%) & 0.7007 $(+24.2\%)$ &  3.7301 (+32.3\%)
 \\ \hline
\end{tabular}
}
\caption{Metrics for predicting the diffusivity $D$ in the test set for exact data using the five most important features. Relative changes (\%) compared to all features (see \autoref{table_metric_exactD}) are provided in parentheses. }\label{table_metric_exactDim}
\end{center}
\end{table}

\autoref{table_metric_exactKim} presents the relative performance for predicting $K.$  All models preserved high $R^2$ scores and kNN displayed a significant relative increase (+73.3\%), suggesting that it has benefited from feature selection. In terms of error metrics, SVM achieved excellent results, maintaining a zero MSE and reducing MAE by 14.3\%. NN also preserved strong accuracy, with virtually no change in MSE and a slight increase in MAE (+4.7\%), indicating stable performance. XGBoost, RF, and kNN all showed notable decreases in both MSE and MAE.

Overall, the results indicate that SVM and NN models retain high accuracy despite using fewer features across both outputs. These findings emphasize the importance of feature selection not only for improving the explainability of the model but also for enhancing computational efficiency.

\begin{table}[t!]
\begin{center}
\scalebox{0.85}{
\begin{tabular}{| c | c  | c  | c  | c  | c| } 
 \hline
Metric & SVM & XGBoost & RF & NN & kNN  
\\ \hline 
$R^2$ score & 1.0000 (0\%) & 0.9792 (+0.8\%) & 0.9690 (+0.8\%) & 0.9999 $(0\%)$ & 0.9758 (+73.3\%)
\\
MSE &   0.0000 (0\%) & 0.0152 $(-28.3\%)$ & 0.0227 $(-20.0\%)$ &  0.0001 (0\%) &  0.0177 $(-22.7\%)$
\\
MAE &  0.0036 $(-14.3\%)$ & 0.1044 $(-14.7\%)$ & 0.1077 $(-10.7\%)$ & 0.0067 (+4.7\%) &  0.0964 $(-13.4\%)$
 \\ \hline
\end{tabular}
}
\caption{Metrics for predicting the conductivity $K$ in the test set for exact data using the five most important features. Relative changes (\%) compared to all features (see \autoref{table_metric_exactK}) are provided in parentheses.}\label{table_metric_exactKim}
\end{center}
\end{table}

\subsubsection{Dimensionality Reduction}

We apply  PCA and UMAP to reduce our 30-dimensional data to two dimensions. This dimensionality reduction is performed exclusively on the training set ($80\%)$ to prevent data leakage. The learned transformations are then applied to the test and validation sets (each $10\%)$.
 
PCA compresses the original set into two orthogonal principal components, each expressed as a weighted linear combination of the initial features. The performance metrics obtained when these components are used as the input features are summarized in \autoref{table_pcaD} and \autoref{table_pcaK}.  For predicting \( D \), all models maintained high \( R^2 \) scores (\( \geq 0.9995 \)), with SVM and NN achieving perfect scores. However, all models experienced an increase in error metrics compared to using all features, most significantly in the MSE and MAE values, with Random Forest and XGBoost showing over 200\% increases in MSE. In contrast, for predicting \( K \), the degradation in \( R^2 \) was more pronounced, especially for XGBoost $(–2.88\%)$ and Random Forest $(–5.54\%),$ indicating a greater sensitivity to feature reduction. Error metrics such as MSE and MAE also rose, with the Neural Network maintaining the most robust performance overall, showing minimal increases in MSE and moderate rises in MAE. These results suggest that while reduced feature sets can still yield accurate predictions for \( D \), recovering \( K \) may require a more comprehensive set of features.

\begin{table}[t!]
\begin{center}
\scalebox{0.8}{
\begin{tabular}{| c | c  | c  | c  | c  | c| } 
 \hline
Metric & SVM & XGBoost & RF & NN & kNN  
\\ \hline 
$R^2$ score & 1.0000 (0\%) & 0.9997 $(-0.02\%)$ & 0.9995 $(-0.04\%)$ & 1.0000 (0\%) & 0.9997 $(-0.01\%)$
\\
MSE & 4.3500 $(+46.0\%)$ & 31.1700 $(+245.8\%)$ & 47.0970 $(+225.5\%)$ & 1.2679 $(+139.5\%)$ & 29.9027 $(+98.3\%)$
\\
MAE & 1.8180 $(+24.3\%)$ & 3.8978 $(+70.4\%)$ & 4.9244 $(+77.0\%)$ & 0.8583 $(+52.2\%)$ & 3.4076 $(+20.8\%)$
 \\ \hline
\end{tabular}
}
\caption{Metrics for predicting the diffusivity $D$ in the test set using PCA generated features. Relative changes (\%) compared to all features (see \autoref{table_metric_exactD}) are provided in parentheses.}
\label{table_pcaD}
\end{center}
\end{table}

\begin{table}[t!]
\begin{center}
\scalebox{0.84}{
\begin{tabular}{| c | c  | c  | c  | c  | c| } 
 \hline
Metric & SVM & XGBoost & RF & NN & kNN  
\\ \hline 
$R^2$ score & 1.0000 (0\%) & 0.9430 $(-2.88\%)$ & 0.9080 $(-5.54\%)$ & 0.9997 $(-0.02\%)$ & 0.9432 $(-2.63\%)$
\\
MSE & 0.0000 (0\%) & 0.0417 $(+96.7\%)$ & 0.0672 $(+136.6\%)$ & 0.0002 $(+100.0\%)$ & 0.0415 $(+81.2\%)$
\\
MAE & 0.0051 $(+21.4\%)$ & 0.1411 $(+35.1\%)$ & 0.1921 $(+59.3\%)$ & 0.0092 $(+43.8\%)$ & 0.1319 $(+18.4\%)$
 \\ \hline
\end{tabular}
}
\caption{Metrics for predicting the conductivity $K$ in the test set using PCA generated features. Relative changes (\%) compared to all features (see \autoref{table_metric_exactK}) are provided in parentheses.}
\label{table_pcaK}
\end{center}
\end{table}

Next, we perform a UMAP reduction, to project our data into a 2D space (to be comparable with PCA). The results in \autoref{table_umapD} highlight the performance degradation of all models when using only the two transformed features from UMAP to predict the diffusivity \( D \). While the \( R^2 \) scores remain relatively high (all above 0.9979), indicating that the models still capture most of the variance, the error metrics increase substantially compared to when all 30 features are used. The NN exhibits the most pronounced increase in MSE $(+25457\%)$ and MAE $(+1368\%),$ suggesting that UMAP-based compression significantly reduces its predictive accuracy. Similarly, the RF and XGBoost models show large MSE increases. Among all models, kNN shows the smallest increase in MSE and MAE, indicating comparatively better robustness under dimensionality reduction. \autoref{table_umapK} shows severe decrease in performance metrics when predicting \( K \) using UMAP. All models exhibit a substantial drop in \( R^2 \) scores, with decreases ranging from approximately 21\% to nearly 40\% compared to using the full set of features. The most noticeable decline is observed for XGBoost and SVM. The error metrics increase dramatically across all models. For instance, the MSE for Random Forest and kNN rises by over 500\%.

Overall, while PCA retains a high-level representation of the data, the loss in predictive precision, particularly in error metrics, suggests that relying solely on the first two principal components may not be sufficient for accurately predicting both parameters. However, the two features generated with UMAP do not capture the essential information required to accurately predict \( K \), highlighting its limitations in this context without further investigation and hyperparameter tuning. 

\begin{table}[t!]
\begin{center}
\scalebox{0.7}{
\begin{tabular}{| c | c  | c  | c  | c  | c| } 
 \hline
Metric & SVM & XGBoost & RF & NN & kNN  
\\ \hline 
$R^2$ score & 0.9990 $(-0.10\%)$ & 0.9983 $(-0.16\%)$ & 0.9979 $(-0.20\%)$ & 0.9986 $(-0.14\%)$ & 0.9991 $(-0.07\%)$
\\
MSE & 98.7650 $(+3216.2\%)$ & 168.2688 $(+1766.3\%)$ & 210.1488 $(+1351.8\%)$ & 135.3956 $(+25457.2\%)$ & 90.5098 $(+500.4\%)$
\\
MAE & 6.9092 $(+372.3\%)$ & 8.8069 $(+285.0\%)$ & 8.8462 $(+218.0\%)$ & 8.2829 $(+1368.7\%)$ & 6.9310 $(+145.8\%)$
 \\ \hline
\end{tabular}
}
\caption{Metrics for predicting the diffusivity $D$ using UMAP features. Relative changes (\%) compared to all features (see \autoref{table_metric_exactD}) are provided in parentheses.}
\label{table_umapD}
\end{center}
\end{table}

\begin{table}[t!]
\begin{center}
\scalebox{0.78}{
\begin{tabular}{| c | c  | c  | c  | c  | c| } 
 \hline
Metric & SVM & XGBoost & RF & NN & kNN  
\\ \hline 
$R^2$ score & 0.6497 $(-35.0\%)$ & 0.5853 $(-39.7\%)$ & 0.7440 $(-22.6\%)$ & 0.6742 $(-32.6\%)$ & 0.7610 $(-21.5\%)$
\\
MSE & 0.2559 $(+\infty\%)$ & 0.3029 $(+1327.8\%)$ & 0.1870 $(+558.5\%)$ & 0.2380 $(+168102\%)$ & 0.1745 $(+662.0\%)$
\\
MAE & 0.3539 $(+8328.6\%)$ & 0.3753 $(+259.4\%)$ & 0.3169 $(+162.9\%)$ & 0.3462 $(+5309.4\%)$ & 0.3065 $(+175.3\%)$
 \\ \hline
\end{tabular}
}
\caption{Metrics for predicting the conductivity $K$ using UMAP features. Relative changes (\%) compared to all features (see \autoref{table_metric_exactK}) are provided in parentheses.}
\label{table_umapK}
\end{center}
\end{table}

\subsection{Noisy data}


Adding noise to simulated data, specifically on the test set, is important when aiming to mimic real-world field infiltration data.  By adding Gaussian noise into the test set, we can better assess how well a model trained on ideal synthetic data generalizes to the noisy conditions encountered in practice. In the following, the noise levels $\delta=2\%$ and $\delta=5\%$ are considered in \eqref{eq_error}.

Adding 2\% noise to the test data significantly affects model performance, particularly for tree-based and ensemble methods. The effect on predicting $D$ is shown in  \autoref{table_metric_relchangeD}. While all models maintain high \( R^2 \) scores, suggesting they still explain most of the variance, their error metrics (MSE and MAE) increase substantially, especially for XGBoost (with MSE increasing by over 9200\% and MAE by over 650\%) and Random Forest.   In contrast, the kNN model shows the smallest relative change in error metrics, suggesting better robustness to noise, despite being a simpler model. The NN maintains an excellent \( R^2 \) and moderate error inflation, demonstrating a balance between accuracy and stability to noise. 

\begin{table}[t!]
\begin{center}
\scalebox{0.75}{\begin{tabular}{| c | c | c | c | c | c |}
\hline
Metric & SVM & XGBoost & RF & NN & kNN \\
\hline
$R^2$ score & 0.9914 $(-0.9\%)$ & 0.9999 (0\%) & 0.9986 $(-0.1\%)$ & 0.9999 $(-0.01\%)$ & 0.9998 (0\%) \\
MSE        & 8.7198 (+192.7\%) & 846.0999 (+9266.6\%) & 139.4403 (+863.1\%) & 12.1855 (+2199.0\%) & 23.6522 (+56.8\%) \\
MAE        & 2.2573 (+54.3\%) & 17.2005 (+651.6\%) & 8.3986 (+201.8\%) & 2.7825 (+393.2\%) & 3.6748 (+30.3\%) \\
\hline
\end{tabular}}
\caption{Performance on predicting $D$ for noisy test data (2\% noise) with relative changes (\%) compared to exact data of \autoref{table_metric_exactD}.}
\label{table_metric_relchangeD}
\end{center}
\end{table}

The presence of noise in the test set degrades model performance also in predicting conductivity \( K \). All models show increased error metrics, with XGBoost and RF particularly impacted. XGBoost's \( R^2 \) drops by nearly $43\%,$ and its MSE and MAE increased by over 1400\% and 2800\%, respectively, signaling a strong sensitivity to noisy inputs, see \autoref{table_reconK}.
The NN and SVM retain relatively high \( R^2 \) scores, indicating they still capture the underlying trend, but their errors increase significantly due to the small-scale nature of the target variable. Interestingly, kNN again shows relatively moderate increases in error, suggesting better noise tolerance. 

\begin{table}[t!]
\begin{center}
\scalebox{0.75}{\begin{tabular}{| c | c | c | c | c | c |}
\hline
Metric & SVM & XGBoost & RF & NN & kNN \\
\hline
$R^2$ score & 0.9864 $(-1.4\%)$ & 0.5541 $(-43.0\%)$ & 0.8008 $(-16.7\%)$ & 0.9756 $(-2.4\%)$ & 0.9546 $(-1.4\%)$ \\
MSE        & 0.0099 (+$\infty$\%) & 0.3257 (+1436.3\%) & 0.1455 (+412.0\%) & 0.0178 (+17700.0\%) & 0.0331 (+44.5\%) \\
MAE        & 0.0757 (+1702.4\%) & 0.4001 (+2832.2\%) & 0.2758 (+1286.4\%) & 0.1093 (+1607.8\%) & 0.1431 (+128.5\%) \\
\hline
\end{tabular}}
\caption{Performance on predicting $K$ for noisy test data (2\% noise), with relative changes (\%) compared to exact data in \autoref{table_metric_exactK}.}
\label{table_metric_relchangeK}
\end{center}
\end{table}

To summarize, introducing noise to the test set significantly affects the prediction accuracy of both soil properties, but with notably different magnitudes and model sensitivities. For diffusivity \( D \), most models maintain high \( R^2 \) values, indicating stable predictive power despite increased error metrics. NN and SVM remain particularly robust, while XGBoost shows the greatest sensitivity.  In contrast, conductivity \( K \) predictions are more severely impacted by noise, especially in terms of \( R^2 \) reduction. XGBoost experiences a drastic performance drop, and error metrics for all models increase by several hundred or oven thousand percent. 

\begin{figure}
\centering
\begin{subfigure}{.5\textwidth}
  \centering
  \includegraphics[height=0.9\textheight]{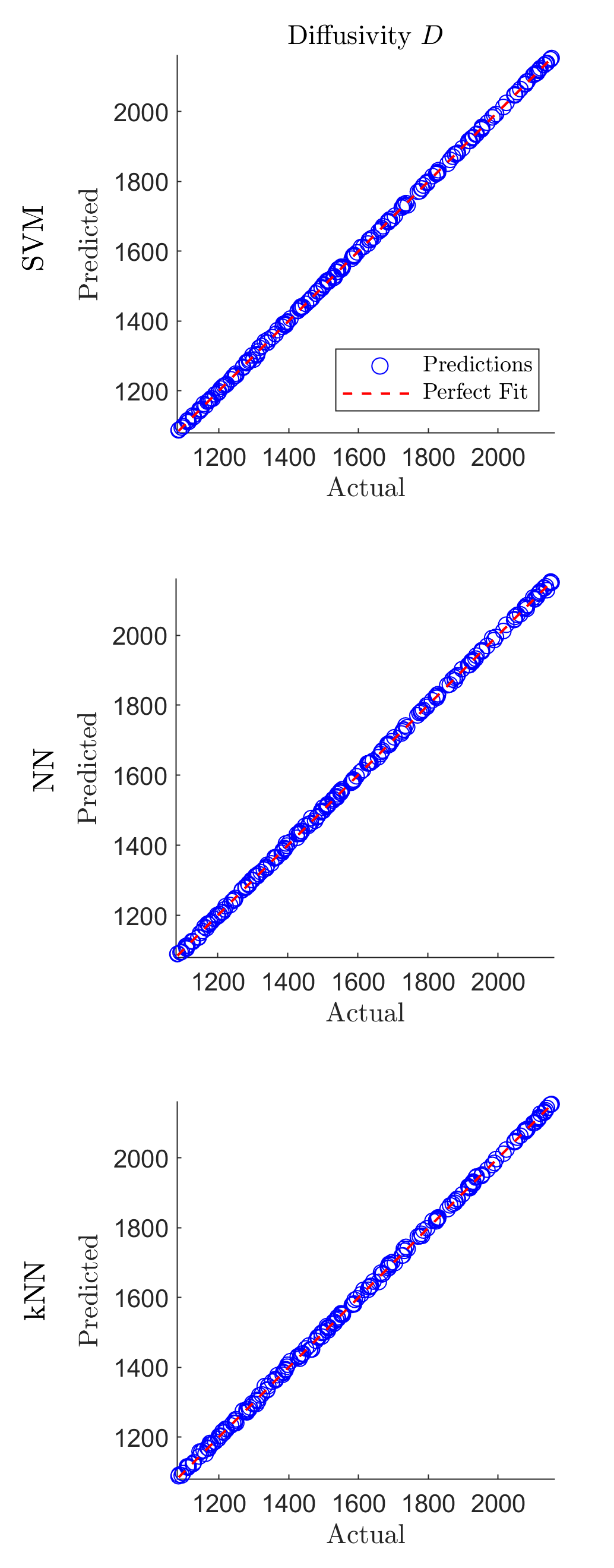}
\end{subfigure}%
\begin{subfigure}{.5\textwidth}
  \centering
  \includegraphics[height=0.9\textheight]{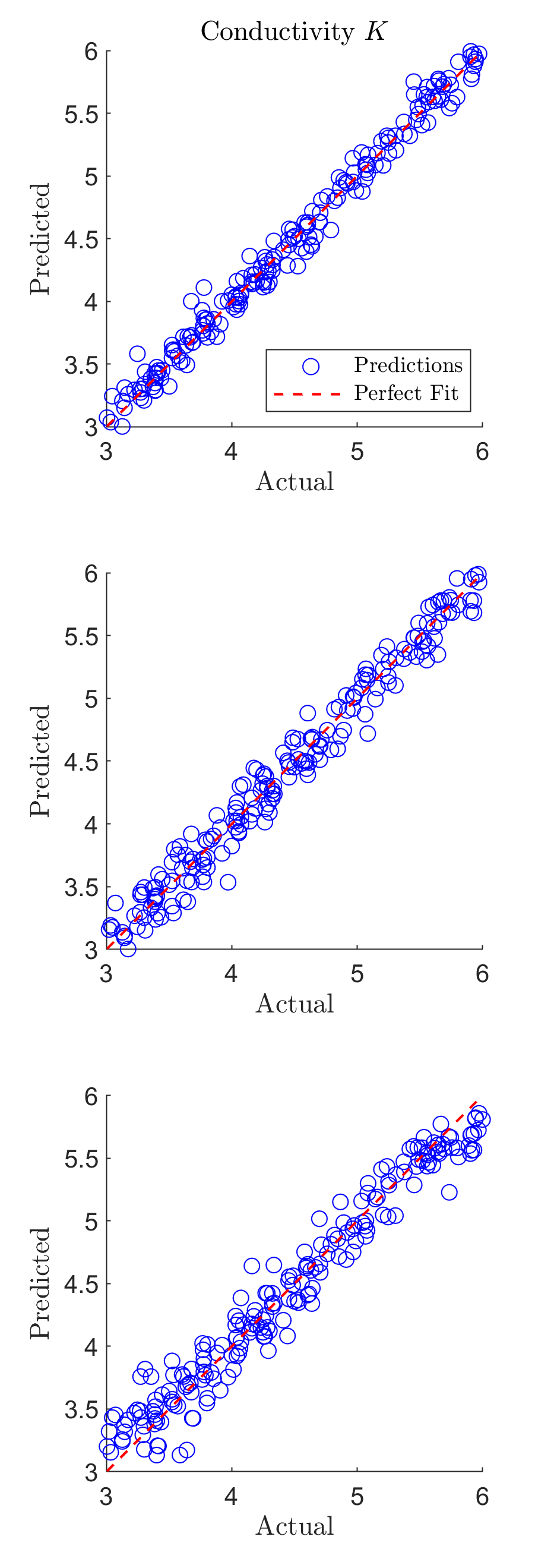}
\end{subfigure}
\caption{Regression plots of $D \,[\SI{}{cm^2/h}]$ (left) and $K \,[\SI{}{cm/h}]$ (right) in the test set for data with $2\%$ noise. Methods that achieved an 
$R^2$ score greater than 0.95 for both outputs are presented.}
\label{fig_regre_noise}
\end{figure}

In \autoref{fig_regre_noise} we present the regression plots of the models SVM, NN and kNN that achieved an $R^2$ score greater than 0.95 on both parameters. The effect of noise is best visualized on predicting $K$ where we observe that most of the scattered points are clustered around the line $y=x$ but not on the line as the pair actual-predicted for $D.$ 

\begin{figure}
\centering
  \includegraphics[height=0.9\textheight]{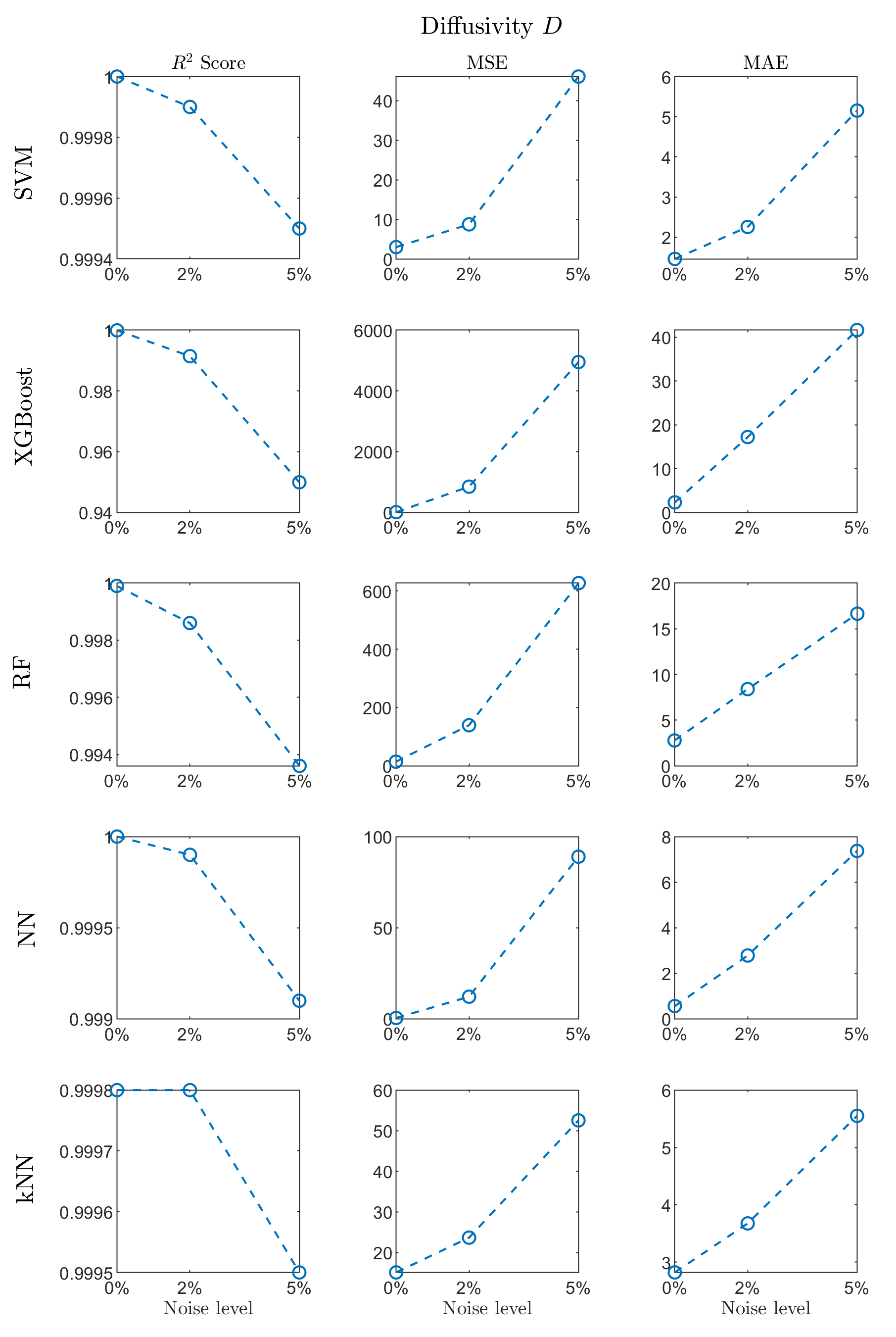}
\caption{The impact of test set noise on the performance metrics for predicting $D$.}
\label{fig_errorD}
\end{figure}

\begin{figure}
\centering
  \includegraphics[height=0.9\textheight]{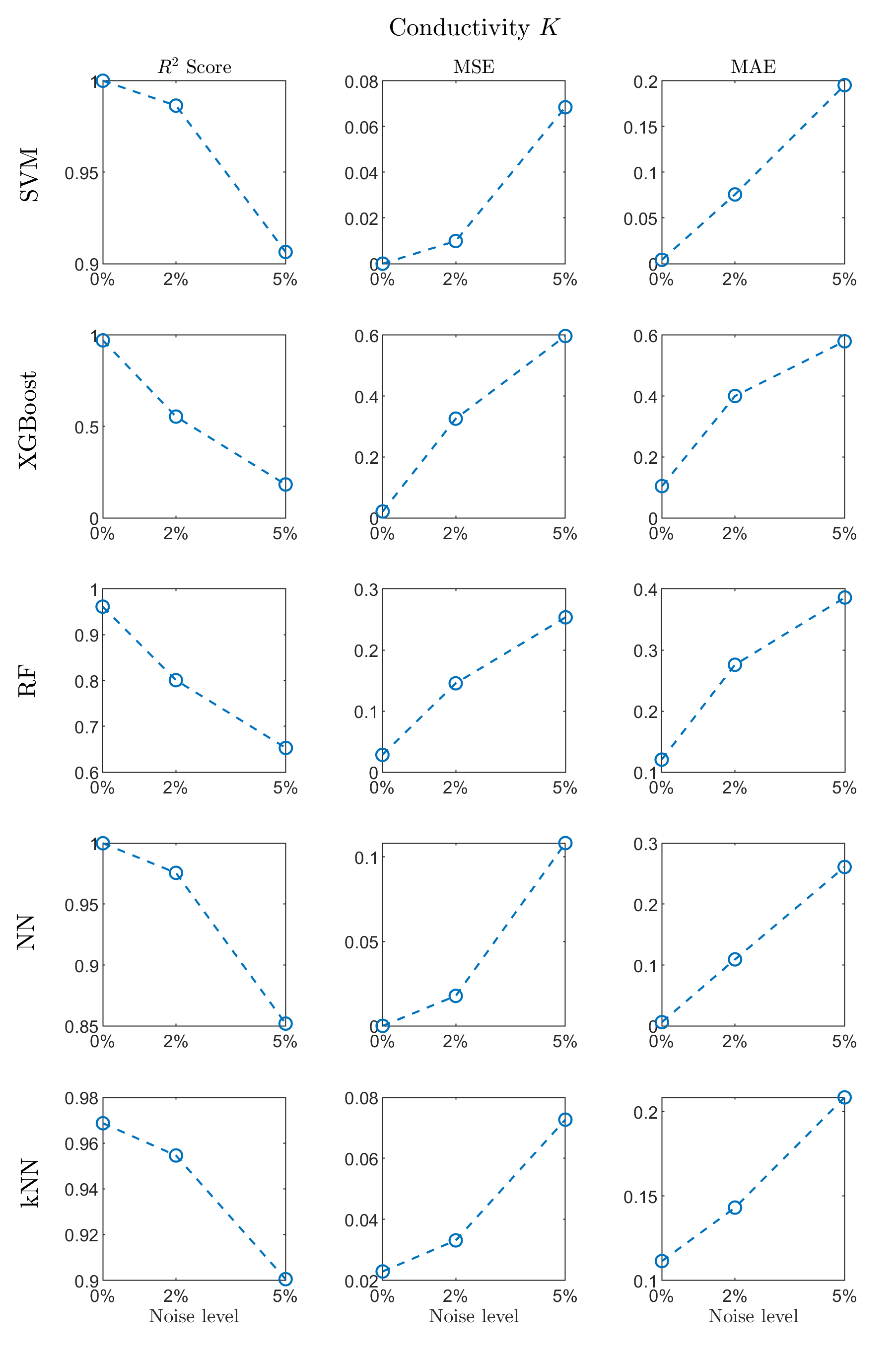}
\caption{The impact of test set noise on the performance metrics for predicting $K$.}
\label{fig_errorK}
\end{figure}

As the noise level increases from $2\%$ to $5\%$, the accuracy $(R^2$ score)  decreases and the error metrics (MSE, MAE) increase, confirming that model predictions become less reliable with noisier data (\autoref{fig_errorD} and \autoref{fig_errorK}). The degree of impact varies across outputs, suggesting that some targets or models are more noise-sensitive than others. More precisely, all models except XGBoost still achieve high $R^2$ score ($> 0.99$) for predicting $D$ with SVM having the lowest MSE and MAE. XGBoost seams to be the least stable model with respect to noise for both outputs. For predicting $K,$ SVM and kNN show similar performance for $5\%$ noise with the best $R^2$ score ($> 0.90$) and the lowest MSE $<0.08$ and MAE $<0.2$ being robust to noise increase. The results emphasize the importance of robustness testing and possibly adding noise during training to improve real-world performance.

\subsection{Limited data}

Reducing the dataset size is a crucial aspect of realistic machine learning applications, especially when thinking of real data. In practical applications, collecting large volumes of data is often expensive and time-consuming. In this section we examine model's performance when the dataset size is reduced mimicking limited data availability, while maintaining the number of (30) features constant. More precisely, we consider three dataset sizes: 2000, 1000 and 500. In general, model performance declines when predicting $K$ as dataset size decreases, meaning that $R^2$ scores drop and MSE and MAE both increase. 

\begin{figure}
\centering
  \includegraphics[height=0.9\textheight]{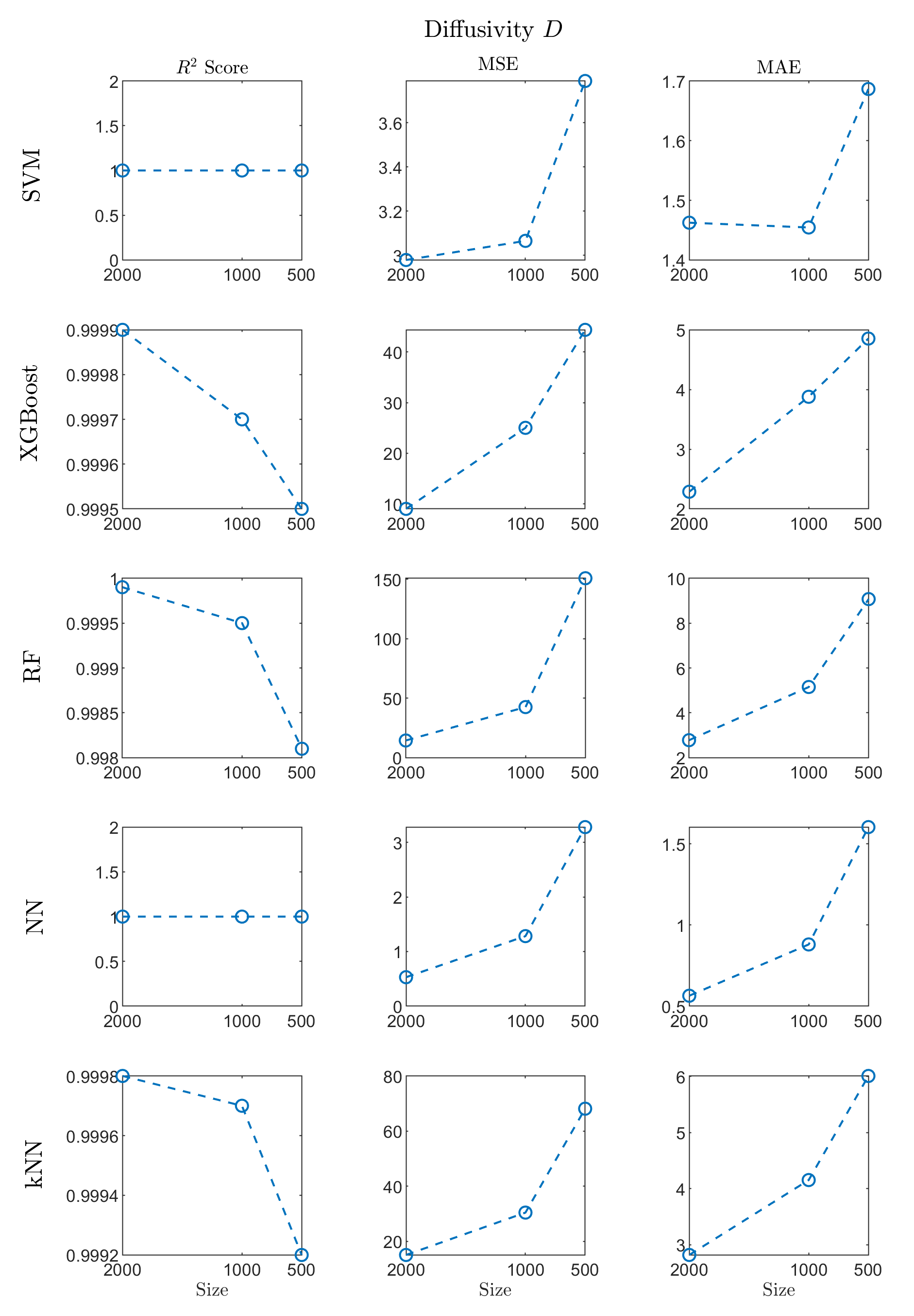}
\caption{The effect of dataset size on the metrics for predicting $D.$ The results refer to the test set for exact data.}
\label{fig_sizeD}
\end{figure}

\begin{figure}
\centering
  \includegraphics[height=0.9\textheight]{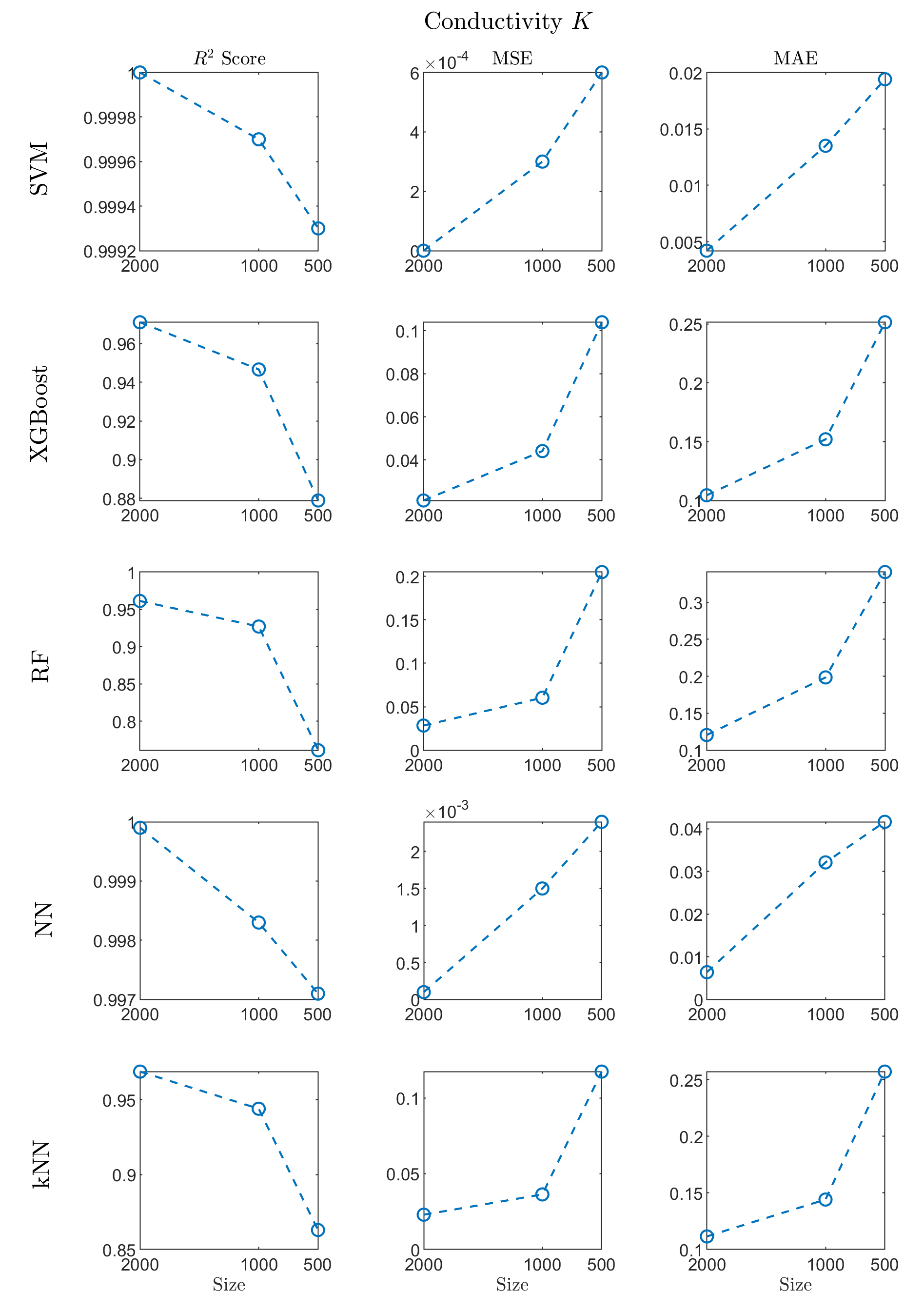}
\caption{The effect of dataset size on the metrics for predicting $K.$ The results refer to the test set for exact data.}
\label{fig_sizeK}
\end{figure}

\begin{figure}
\centering
  \includegraphics[height=0.9\textheight]{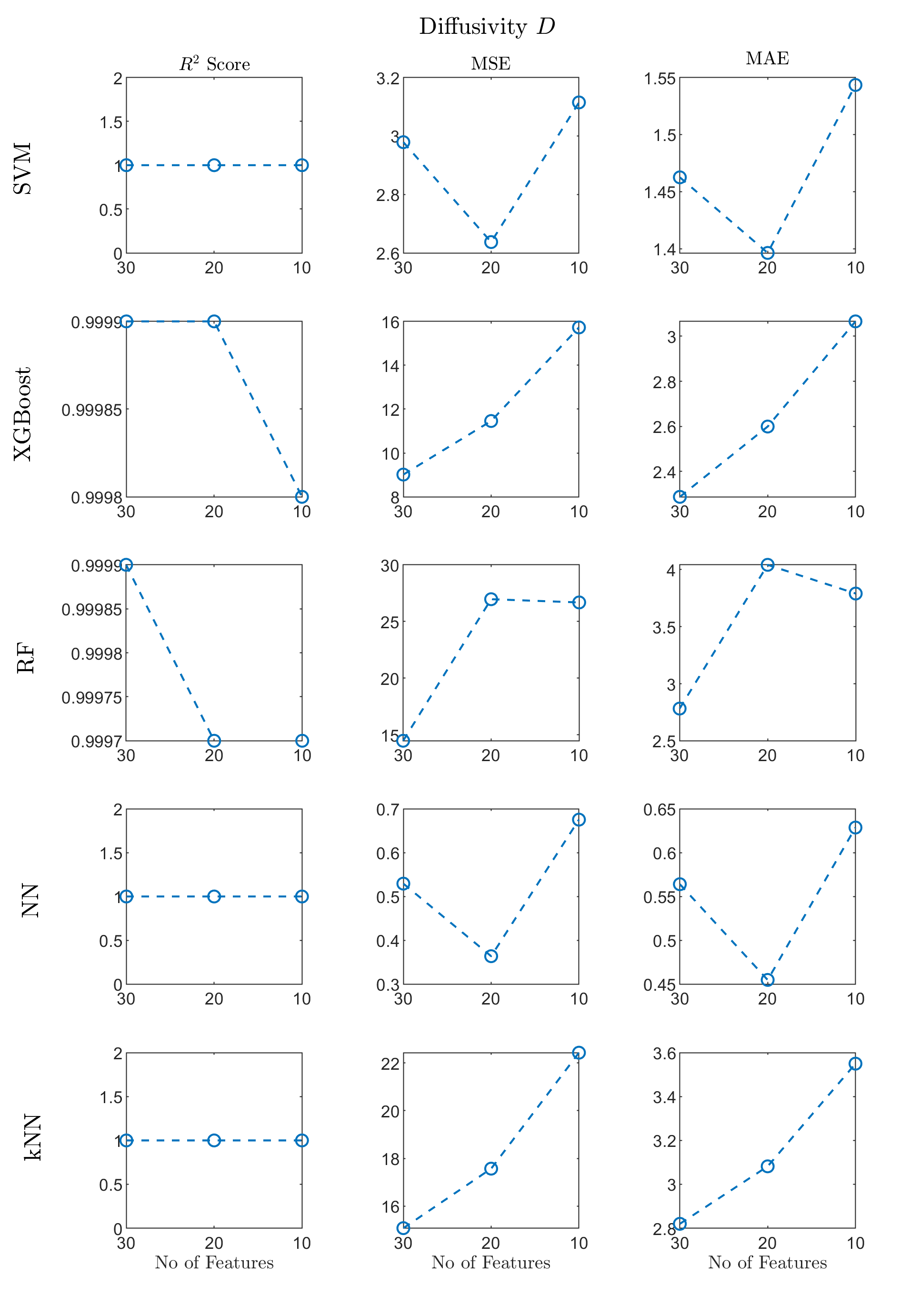}
\caption{The effect of feature count on the metrics for predicting $D.$ The results refer to the test set for exact data.}
\label{fig_featD}
\end{figure}

\begin{figure}
\centering
  \includegraphics[height=0.9\textheight]{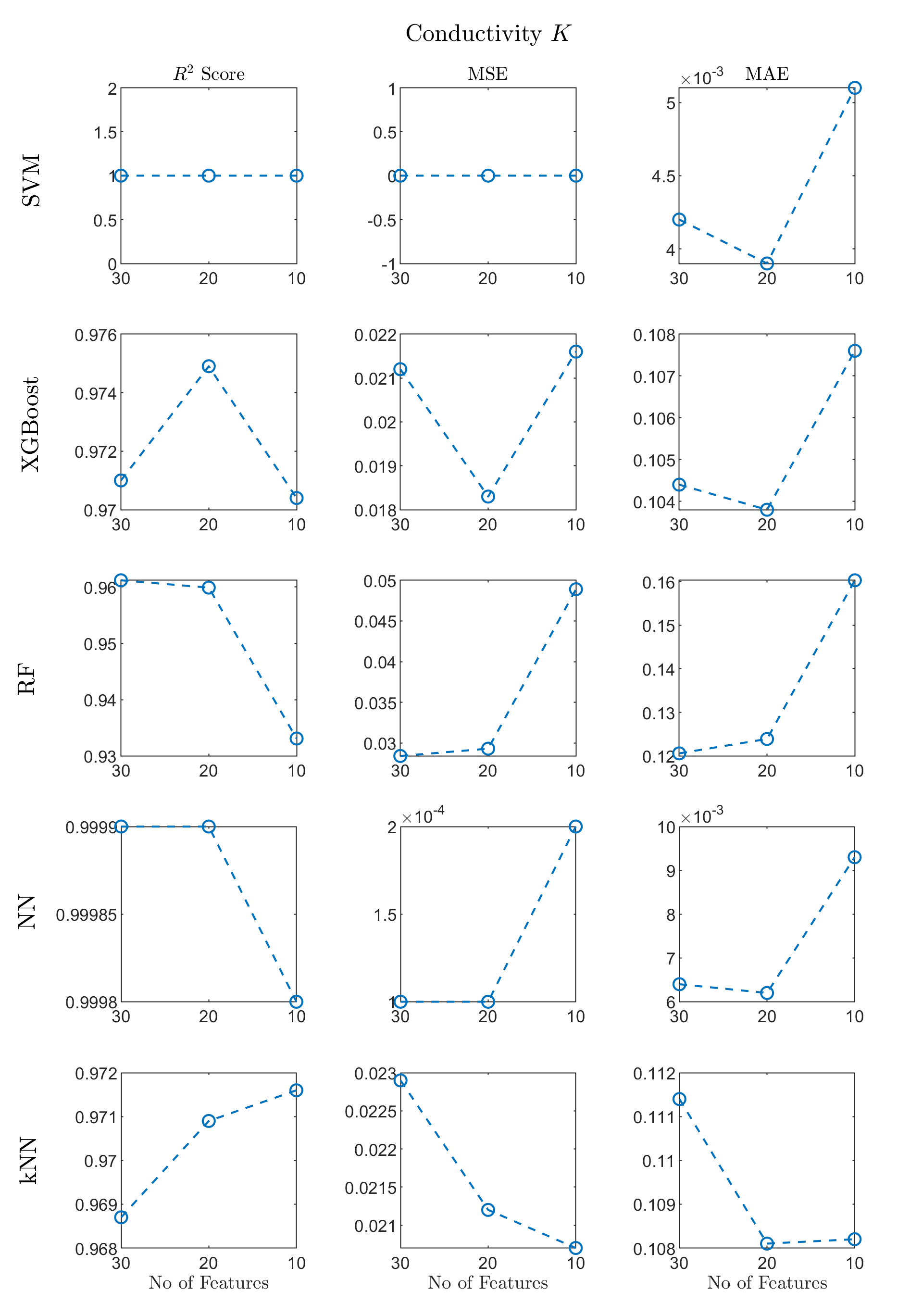}
\caption{The effect of feature count on the metrics for predicting $K.$ The results refer to the test set for exact data.}
\label{fig_featK}
\end{figure}

From \autoref{fig_sizeD} and \autoref{fig_sizeK}, it is evident that SVM and NN are the most robust models to dataset size reduction, maintaining high $R^2$ values (above 0.99) even with as few as 500 samples. Their error metrics (MSE and MAE) increase more gradually compared to the tree-based models. In contrast, XGBoost and RF show greater sensitivity to reduced dataset size, with RF exhibiting the highest MSE and MAE and the lowest $R^2$ score when predicting 
$K.$ kNN displays moderate sensitivity to dataset size but still outperforms RF under smaller data conditions.

In summary, decreasing dataset size influences model performance when predicting $K$, with varying effects across different machine learning algorithms. Notably, SVM and NN stand out as the most robust models, maintaining near-optimal accuracy even under reduced data.

\section*{Conclusions}

We investigated the inverse problem of estimating soil properties using various machine learning (ML) models.  The study focused on vertical infiltration under flooding conditions, where the water content is measured at different depths and times. To train, test and evaluate these models, we generated both exact and noisy simulated data by numerically evaluating the analytical solution of the direct problem, as derived using the Fokas method.

In summary, the evaluation of ML models on the full dataset reveals that Support Vector Machines (SVM) and Neural Networks (NN) consistently outperform other models in predicting both diffusivity ($D$) and conductivity ($K$). These models achieved perfect or near-perfect $R^2$ scores, minimal mean squared error (MSE), and low mean absolute error (MAE), with NN slightly outperforming SVM in terms of precision. XGBoost demonstrated solid but less consistent accuracy, while Random Forests (RF) and k-Nearest Neighbors (kNN) exhibited higher error metrics, particularly in estimating diffusivity. Cross-validation results confirmed these trends, highlighting the robustness and generalization capability of SVM and NN.

The feature analysis emphasized the significant role of a few key measurements, particularly those near the surface and at earlier time steps, in driving model predictions. 
Using only the measurements from the top data point (10 features) still yielded strong performance. Notably, models maintained high accuracy even when restricted to only the five most important features, as determined by feature importance rankings. On the other hand, the application of dimensionality reduction techniques (PCA and UMAP) resulted into reduced performance across all models.

Regarding the impact of noise and dataset size  on model performance, we found that these factors significantly influence the results, with varying effects across different machine learning algorithms. Notably, SVM and NN stand out as the most stable models, consistently achieving near-optimal accuracy even with limited data. These findings further establish SVM and NN as the most reliable models for this inverse modeling task.

\paragraph*{Conflict of Interest:} The authors declare no conflict of interest.

\paragraph*{Funding:} KK acknowledges support by the Sectoral Development Program (SDP 5223471) of the Ministry of Education, Religious Affairs and Sports, through the National Development Program (NDP) 2021-25, grant no 200/1029.

\paragraph*{Author Contributions:}  Conceptualization: K.K. and L.M.; Methodology: L.M. and N.P.; Software: L.M. and N.P.; Validation: K.K., L.M. and N.P.; Formal Analysis: K.K. and L.M.; Investigation: K.K., L.M. and N.P.; Data Curation: L.M. and N.P.;  Writing – Review \& Editing: K.K., L.M. and N.P.; Visualization: L.M.;

\bibliographystyle{siam}
\bibliography{refs}

\end{document}